\documentclass[12pt,preprint]{aastex}
\newcommand{\kms}{km s$^{-1}$\hspace{0.25em}}
\newcommand{\etal}{et al.\,~}
\slugcomment{Submitted to Astrophysical Journal}
\shorttitle{VLBI Water Maser Observations of IRAS4A/B}
\shortauthors{K.B. Marvel, \etal}
\begin{document}
\title{Time-Resolved AU-Scale Jets Traced by Masers in the IRAS~4A/B Regions of NGC1333}
\author{Kevin B. Marvel}
\affil{American Astronomical Society}
\affil{2000 Florida Avenue, NW, Suite 400, Washington, DC 20009}
\email{marvel@aas.org}
\author{Bruce A. Wilking}
\affil{Department of Physics and Astronomy}
\affil{University of Missouri at St. Louis, 1 University Boulevard, St. Louis, MO 63121}
\email{bwilking@umsl.edu}
\author{Mark J. Claussen}
\affil{National Radio Astronomy Observatory (NRAO) Array Operations Center}
\affil{P.O. Box 0}
\affil{1003 Lopezville Road, Socorro, NM 87801}
\email{mclausse@nrao.edu}
\author{Alwyn Wootten}
\affil{NRAO}
\affil{520 Edgemont Road, Charlottesville, VA 22903-2475}
\email{awootten@nrao.edu}
\begin{abstract}
We present results of VLBI observations of the water masers associated 
with IRAS 4A and IRAS 4B in the NGC 1333 star-forming region taken in four
epochs over a two month period.
Both objects have 
been classified as extremely young sources and each source is known to be a multiple system.   
Using the Very Long Baseline Array, we detected 35 masers in Epoch I, 40 masers in Epoch II, 
35 in Epoch III, and 24 in Epoch IV.  Only one identified source in each system associates with these masers.  These data are used to calculate proper motions for the masers and trace the jet outflows within 100 AU of IRAS 4A2 and IRAS 4BW.  
In IRAS 4A2 there are two groups of masers, one near the systemic cloud velocity 
and one red-shifted. 
They expand linearly away from each other at velocities of 53 \kms.
In IRAS 4BW, masers are observed in two groups that are blue-shifted and red-shifted
relative to the cloud velocity.  They form complex linear structures with a thickness of 3 mas 
(1 AU at a distance of 320 pc)
that expand linearly away from each other at velocities of 78 \kms.
Neither of the jet outflows traced by the maser groups align with the larger scale outflows.
We suggest the presence of unresolved companions to both IRAS 4A2 and 4BW.
\end{abstract}
\section{Introduction}

Water masers are excellent probes of astrophysical flows. Their large flux
densities and exceedingly compact sizes make them perfect Very Long Baseline
Interferometer (VLBI) targets.  
Many masers observed in star-forming regions are thought to form in shocks 
either in or along the outflow of material commonly seen associated with 
young stellar objects (YSOs).  They provide a high resolution probe of the 
base of the stellar wind that is unaffected by extinction from dust
or extensive interaction with ambient material.  
The milliarcsecond resolution of VLBI is critical to resolve the individual 
maser components and outflows, particularly in high stellar density environments. 

Long term total power monitoring has shown that masers around low mass YSOs 
are more episodic than those around higher mass YSOs, however, their 
detectable phases are sufficiently long for proper motion measurements if the observations 
are spaced by no more than two to three weeks \citep{Furuya:2003,Brand:2003,
CLAUSSEN:1996,WILKING:1994}.  Furthermore, if they arise in the warm (400K) 
spatially confined post-shock region, as postulated in various models 
\citep{MODEL1, MODEL2, MODEL3, MODEL4}, they should have space velocities 
sufficient to produce measurable proper motions over one to three weeks.  
However, regular monitoring of low mass YSOs is necessary to ensure that 
VLBI observations are conducted during periods of high maser activity. 

Using the NRAO's \footnote{The National Radio Astronomy Observatory is a facility 
of the National Science Foundation operated under cooperative agreement 
by Associated Universities, Inc.} 
Very Long Baseline Array (VLBA), proper motion studies have been 
made of water masers associated with high mass and intermediate mass YSOs 
such as Cepheus A HW2 \citep{TORRELLES:2001}, S~106 FIR \citep{FURUYA:2000},  NGC 2071 \citep{SETH:2002}, and IRAS~20050+2720 MMS1 \citep{FURUYA:2005} as well as with lower mass YSOs such as Serpens SMM1, RNO 15-FIR, and IRAS~05413-0104 \citep{Moscadelli:2006,CLAUSSEN:1998}.  
For example, water masers in IRAS~05413-0104 (associated with HH~212 at
a distance of 450 pc), were 
found to lie in a structure of 10 AU in length and less than 0.5 AU thick. 
The masers appeared to arise in shock-related structures which showed proper motions 
along the axis of the outflow of 60 \kms and displayed coherent structures 
over time scales of two to three weeks (but which was less discernible 
over time scales as long as several months).  Comparison with molecular 
and infrared observations of IRAS~05413-0104 clearly demonstrated that 
proper motion studies of water masers in the region (within 40 AU) of the 
central source are among the best tools for studying the kinematics 
of the jets emanating from embedded YSOs.  

IRAS~4, comprised of multiple sources whose submillimeter-dominated 'Class 0' spectral energy distributions suggest extreme youth, is located in the NGC 1333 star-forming 
region.  As a nearby active and multiple young object,  it presents a prime candidate for VLBI observations of its associated water masers. 
Located at a distance of 320 pc \citep{HIPPARCOS}, the NGC 1333 molecular 
cloud hosts a double infrared cluster of about 200 YSOs identified in 
near-infrared surveys \citep{STROM:1976,ASPIN:1994,Lada:1996,Wilking:2004} 
and x-ray surveys \citep{PREIBISCH:1997,Getman:2002,PREIBISCH:2003}.  
Far-infrared observations have not only revealed the higher luminosity 
sources in NGC 1333, but also YSOs in the earliest phase of evolution 
such as the Class 0 sources IRAS~4 and IRAS~2 \citep{Harvey:1984,JENNINGS:1987}.
IRAS~4 was found to be a binary at submillimeter wavelengths \citep{SANDELL:1991};
hence IRAS~4A and 4B.  IRAS~4A is located 31$\arcsec$ to the 
NW of 4B at a position angle of -45$^{\circ}$.
Subsequently, both IRAS~4A and 4B were found to be binary systems.
\citet{LAY:1995} performed single baseline interferometry in the submillimeter 
and found that 4A was a binary with an angular separation of 1.8$\arcsec$
(or 580 AU at 320 pc).  Interferometric observations in the millimeter 
and radio continuum confirmed the separation and determined that the 
position angle of separation was $\approx -20^{\circ}$ \citep{LOONEY:2000,REIPURTH:2002}.  
While the easternmost source (IRAS~4A1) dominates the millimeter and radio continuum emission, 
the westernmost source IRAS~4A2) appears to be relatively more evolved with a warm 
ammonia core \citep{WOOTTEN:1993,Shah:2000} and a wealth of complex organic molecules 
(Bottinellii 2007, priv. comm.).  Submillimeter and millimeter wave continuum observations 
have confirmed the prediction of \citet{LAY:1995} that IRAS~4B is also a 
binary with components (4BE and 4BW) separated by about 10$\arcsec$ (3200 AU in 
projection) along an E-W direction \citep{SMITH:2000,LOONEY:2000,Sandell:2001}.
\footnote{We adopt the naming convention introduced by \citet{Sandell:2001}.  
The western (eastern) source is referred to as IRAS~4BI (IRAS~4BII) by 
\citet{SMITH:2000}, as IRAS~4B (IRAS~4C) by \citet{LOONEY:2000}, and as 
IRAS~4B (4B$\arcmin$) by \citep{DIFRANCESCO:2001}.}  As in the case of IRAS~4A2, 
complex organic molecules have been detected toward IRAS~4B \citep{Bottinelli:2007}, 
presumably from a hot core associated with 4BW.
We note that IRAS~4BE has yet to be detected in the radio continuum \citep{REIPURTH:2002}
or demonstrate any sign of outflow activity.  Infall motions have been detected toward IRAS~4A and 4B
\citep{DIFRANCESCO:2001}, unambiguously toward the former, and strongly indicated toward the latter.

The detection of water masers toward IRAS~4A and 4B suggests that they are YSOs 
also associated with mass outflow.  Water masers were first detected in single 
dish observations by \citet{HASCHICK:1980} and later monitored by 
\citet{CLAUSSEN:1996}.  These observations established the maser emission 
as highly variable on monthly time scales, sometimes being completely 
absent while at other times reaching peaks of 10 Jy.  Detected emission 
was found from $-$10 to $+$15 \kms but the angular resolution was not sufficient 
to determine which masers were associated with 4A and 4B.  Beginning in 1983, 
VLA observations have shown water maser activity associated only with IRAS~4A2 
and/or 4BW \citep{RODRIGUEZ:2002,Wootten:1998,Furuya:2003}.  VLBA observations 
of the water masers in IRAS~4A and 4B, acquired in 2003, have been presented 
by \citet{Desmurs:2006}.  While masers were detected in only two of the four 
epochs observed, they confirmed the association of masers with IRAS~4A2 
and expansion motion between the two spots was detected.  
In IRAS~4BW, the five maser spots detected were red-shifted and formed a chain 
80 mas in extent with some masers displaying proper motion toward the north.

Molecular outflows have been mapped toward both IRAS~4A and 4B. 
IRAS~4A is associated with a highly collimated outflow about 20,000 AU 
in extent seen in HCN and SiO with a position angle close to 20$^{\circ}$
\citep{Girart:1999,Choi:2001,Choi:2005}.  The origin of the outflow is likely 
IRAS~4A2 \citep{Choi:2005}.  On a larger scale ($\approx 1 \times 10^{5}$ AU), 
the outflow defined by CO and molecular hydrogen has a position angle of 
45$^{\circ}$ \citep{BLAKE:1995,Choi:2006}.  The shift in position angle from 
small to large scale is perhaps due to a combination of a shift in the direction 
of the magnetic field, an encounter with denser ambient gas, and precession of 
the outflow axis \citep{Choi:2006}.  A more compact outflow is associated with 
IRAS~4BW, with a position angle close to 0$^{\circ}$ \citep{BLAKE:1995,Choi:2001} 
with perhaps a second outflow with a position angle of -35$^{\circ}$ \citep{DIFRANCESCO:2001}.

In this paper we present VLBA observations of water masers associated with the
IRAS~4 region obtained over four epochs in 1998 and spaced by about one month.  
We also describe Very Large Array (VLA)  observations, taken one month in advance of 
the start of the VLBA observations, that showed IRAS~4 to be a very active 
maser source and yielded absolute positions for the masers and the 1.3 cm 
continuum sources IRAS~4A1 and IRAS~4BW.  With the VLBA, we detected 35 masers in 
Epoch I, 40 masers in Epoch II, 35 in Epoch III, and 24 in Epoch IV 
associated with IRAS~4A2 and IRAS~4BW.  We use these data to reveal 
the structure of shock fronts associated with stellar winds from these YSOs 
and to estimate proper motions for masers detected in all four epochs.  
The origin of the maser emission and its relationship to the stellar winds 
and larger scale molecular outflows from these YSOs is discussed.
  
\section{Observations and Reductions}

\subsection{VLA Observations}

The VLA was used on August 12, 1998 to observe the water masers
toward IRAS~4A and IRAS~4B to check that the masers were strong
enough to observe subsequently with the VLBA.  The array was in the
{\bf B} configuration.  The phase tracking center for the array was
set to be midway between the IRAS~4A and IRAS~4B positions.
spread over a range of hour angles to optimize the {\it u, v} coverage.
Two observation modes were used.  In the first, observations were made
with both circular polarizations, a total bandwidth of 1.5625 MHz
(total velocity extent of 21 \kms) and a correlator setup which gave
127 spectral channels each with width of 12.2 kHz (0.16 \kms).  In this
mode the water masers were observed with reasonably high spectral
resolution in order to spectrally resolve the masers, and
the observations lasted for $\sim$42 minutes (on source),

In the second mode, the correlator was configured for continuum
observations with one polarization pair (of IFs) set to a center
frequency of 22264.9 MHz and a bandwidth of 50 MHz, while the other
polarization pair were set to a bandwidth of 0.781 kHz, and the
center frequency adjusted so that the strongest masers were placed
at the center of the narrow band.  This setup allowed the phases
to be tracked using the strong signal in the narrow band, in order
to calibrate tropospheric phase changes to apply to the 50 MHz
continuum band (away from the maser emission). In this mode the
continuum emission from the IRAS~4 region was observed for $\sim$3.5
hours (on source).

The data were edited, calibrated, and imaged with the NRAO Astronomical
Image Processing System (AIPS, Greisen 2003) in the standard manner.  In the high
spectral resolution data, Stokes {\bf I} images were made using uniform weighting
of the {\it u,v} data, of each spectral channel, and were CLEANed
to an rms noise (in channels without strong maser features) of $\sim$20 mJy beam$^{-1}$.
The synthesized beam was 300 by 275 milliarcseconds (mas) at a position angle
of $-$83$^{\circ}$.  The strong masers in SSV13, some 196 arcseconds to the
north-northwest, contributed sidelobe emission in some spectral channels.
In general, however, those channels affected did not have maser emission
from IRAS~4A or 4B.  With this angular resolution, many spectral features
mapped to the same position, although two groups each of spatially separated
masers were seen toward both IRAS~4A and 4B.  The positions of these
four maser groups, as estimated with Gaussian fits to the strongest
maser in each group using the AIPS task JMFIT, are listed in Table 1.

For the continuum observations, after the standard calibration, the narrow
band data were imaged and used in an iterative self-calibration procedure.
The self-calibration solutions obtained were applied to the broad band
data, which were imaged with natural weighting of the {\it u,v} data
(which increases the sensitivity relative to uniform weighting).  The
synthesized beam  for this mode was 360 by 325 mas at a position angle
of 77$\deg$.  The rms noise in this image was $\sim$120 $\mu$Jy beam$^{-1}$.
Unfortunately, the weather was not particularly good, and the system temperatures
were elevated by about a factor 1.5 - 2.0 over good weather conditions. Integrated flux densities 
at $\lambda$=1.3 cm 
for IRAS~4A1 and 4BW were 1.33 mJy and 0.54 mJy, respectively.
IRAS~4A2 and 4BE were not detected to a 3$\sigma$ peak flux density limit of 0.36 mJy beam$^{-1}$.

\subsection{VLBA Observations}

We have used the VLBA to obtain four epochs of observation
of the water maser emission associated with a number of low-mass YSOs.  
Two subsequent papers will present results on IRAS~16293-2422 and SVS 13.
A previous paper \citep{CLAUSSEN:1998} presented results for IRAS~05413-0104.
Preliminary results of the observations presented here have appeared as \citet{MARVEL:2002}, 
while those of IRAS~16293-2422 and SVS 13 have appeared as \citet{WOOTTEN:1999}
and \citet{WOOTTEN:2002}.

The observations were made with one polarization pair (both right and left circular
polarization) of baseband channels, each with 4 MHz of bandwidth.  This provides
54 \kms of total velocity coverage.  Correlation was made with 256 spectral channels
per baseband channel, yielding a velocity width in each spectral channel of 0.21 \kms.
Although the velocity coverage is then only $\pm$27 \kms from the systemic, no 
published higher velocity water masers have been detected from the IRAS~4 sources
\citep{CLAUSSEN:1996, Brand:2003, Furuya:2003}.

Since IRAS~4A and 4B are located close together on the sky (within the primary beam of the
VLBA antennas), we were able to direct
the antennas of the VLBA to a position between the two sources and then perform
two separate correlation passes with the VLBA correlator at the precise
positions of the sources without loss of sensitivity.  
The correlation positions coincide with the centimeter continuum positions
(see Table 1). 
Since the data were obtained contemporaneously, simply
being correlated in separate passes, the two data sets share 
identical time ranges and mid-times of observation.
Each of the four observations were separated in time by approximately three weeks.  
This separation was chosen based on past single dish monitoring observations 
\citep{CLAUSSEN:1996, WILKING:1994}, which indicated significant flux 
changes on this timescale.  The dates of the observations, mid-times 
of each epoch and Julian date of the mid-times are 
given in Table 2 along with the number of days
between each pair of the observations.

The data from each epoch were reduced using AIPS.  Strong
continuum sources were observed during each epoch to calibrate both the 
bandpass response of the receiver systems of each antenna as well as allowing
determination of any residual delays uncompensated for by the correlator
model.  After residual delay removal, a strong maser feature displaying 
simple structure was selected for determination of delay changes during the
observations.   After the removal of these residual delay rates, the same
channel was split from the data set for iterative self-calibration.  The
amplitude and phase self-calibration solutions were then applied to all
spectral channels and the data were edited for poorly calibrated amplitudes.

After all calibrations were applied, the data were mapped using the AIPS task
IMAGR.  Spectral cubes were formed with pixel cell sizes of 85 $\mu$arcseconds.
The restoring beam varied in size, but was typically about 800 $\mu$arcseconds
by 400 $\mu$arcseconds with a position angle very close to 
$-$10.0$^{\circ}$  east of north.  The resolution varied for each epoch
depending on the details of the self-calibration process and whether or not
good self-calibration solutions were obtained for given antennas at given
times.  The exact imaging parameters for each epoch 
and rms noises for typical channels (those channels whose noise floor is
not dictated by dynamic range constraints) are listed in Table 3.

To aid in the analysis of the maser motions observed, two-dimensional Gaussian
components were fit to each maser spot in each channel where the emission
from an individual spot exceeded five times the RMS noise for that channel.
This was carried out using an automated routine in AIPS (SAD) in an iterative
process that first fit the strongest Gaussians in a given channel, subtracted
the resulting component from the map and then reiterated the search for Gaussian
components above the noise cutoff in the map.  This produces a fairly reliable
set of Gaussian components without the very tedious process of fitting each component
above the limit in each channel manually.
The resulting positions were then grouped together in velocity space and position 
in tabular form.
Only masers detected in two or more channels above the cutoff and no more than one
beamwidth offset in position were retained.  Fewer than 10\% of the components were excluded
in this process, pointing to the reliability of the iterative fitting process. 

The surviving component fits were then visually inspected using the 3-dimensional data analysis
tool XGOBI\footnote{XGOBI, a multivariate data analysis tool that operates under the X11 system
is freely available for download at http://www.research.att.com/areas/stat/xgobi/}.
Using XGOBI, it is possible to excise spurious Gaussian fits that result from
the automated Gaussian fitting routine we utilized.  If a fit was
called into question by the XGOBI inspection process, the original maps were inspected
prior to deleting the data point.  Usually, spurious fits were caused by residual sidelobes
being fit in several channels by the SAD routine in AIPS and occurred in channels that were dynamic-range
limited.  The components were then used for all subsequent fitting and analysis as described in Appendix A.

\section{Description and Analysis of the Maser Regions\label{descriptions}}

\subsection{IRAS~4A\label{descriptions-4a}}

At the time of our observations, the water masers of IRAS~4A were 
found in two dominant regions, separated by 251 milliarcseconds (Epoch I) 
or 80.3 AU.  A line connecting the two maser regions lies at a position 
angle of about $-$50$^{\circ}$ (north through east)
roughly  
perpendicular to the CO outflow axis.  Comparison with our VLA observations
show that the masers are associated with IRAS~4A2.
Figure 1 shows
the spatial and kinematic distribution of the masers for all four epochs, 
each with a closeup of the SE maser component.

The NW region exhibited only a single maser spot.  Its peak flux decreased 
in time from 0.97 Jy at the first epoch to 0.12 Jy at the final epoch. 
The maser is at or slightly red-shifted relative to the molecular core velocity of $+$6.7 \kms. 
The SE region contains more highly red-shifted maser spots in Epochs I and II, 
including the 
maser used to self-calibrate each epoch and another weaker maser 
just west of the reference spot.  In Epochs III and IV, another maser 
appeared $\sim$10 mas north of the reference feature at a velocity of $\sim$12.5 \kms (Figure 1).  
Because we self-calibrated the data using the brightest spot as the reference, 
its position remains constant over time and appears at the origin.

The separation vector from the reference feature to the NW feature 
increased in magnitude over time, expanding from 80.4 AU in the first 
epoch to 82.3 AU in the final epoch.  
The projected speed of expansion is 53.0 \kms given the assumed distance of 
320 pc to the source.  Since the line of sight velocity difference between 
the two regions is only 2 \kms and assuming that this velocity difference 
and the proper motion result from the projected velocity of an outflow jet, 
we can calculate an inclination for the outflow of 2$^{\circ}$.  The true 
space velocity is therefore nearly equal to that measured in the plane of the 
sky, namely 53 \kms.  As shown in Table 4, 
the proper motion vectors for the NW feature relative 
to the reference feature has a position angle of $-$49.8$^{\circ}$, identical to
the position angle of the separation vector.

\subsection{IRAS~4B\label{descriptions-4b}}

Similar to IRAS~4A, IRAS~4B exhibits water maser emission in two dominant 
well-separated regions associated with IRAS~4BW: one to the NW and another to the SE.  
Masers in the NW region are red-shifted relative to the cloud velocity of 
7 \kms with velocities ranging from 13 to 20 \kms.  Masers in the SE regions 
are blue-shifted with velocities ranging from -3 to 4 \kms.  The radial 
velocities for the NW and SE groups are consistent with those in the 
northern and southern lobes of the bipolar HCN outflow observed by \citet{Choi:2001} 
and the bipolar H$_2$CO outflow mapped by \citet{DIFRANCESCO:2001}.
A complete list of the maser component velocities, fluxes,
and positions can be found in Appendix A.

Figures 2 and 3 show the overall spatial and kinematic 
distribution of the maser emission as observed in each epoch for the two maser groups in IRAS~4B.  
Over time, the masers in the NW expand away from the reference feature at (0,0) 
in the SE maser group.  A line connecting the two regions lies at a position 
angle of roughly $-$29$^{\circ}$ and the two regions are separated 
by 488 milliarcseconds (mas) or 156 AU.  The SE grouping is 
roughly arc-like with an overall extent of 38 mas (12 AU) and a 
thickness of 2 mas (0.6 AU), while the NW clump is more linear 
and extends about 26 mas (8.3 AU) with a thickness roughly the 
same as the SE clump.  The highest velocity maser emission lies at the 
northern tip of the NW group (+20 \kms) and at the southern 
tip of the SE arc ($-$3 \kms).  In addition, a small maser group 
is located approximately 100 mas (32 AU) to the NNW of the main SE arc. 
This region's structure changed during the epochs, but usually consisted of 
two or three well-separated maser features (by several mas or about 0.3 AU).  
This distinct region is located close to the line connecting the 
NW and SW maser groups.  If the active maser regions mark the interaction of a 
bipolar jet from IRAS~4BW with ambient gas, then the opening angle of the jet 
would be $\sim$10$^{\circ}$.

Beginning in the second epoch, but much more clearly shown in the third 
and fourth epochs, a new arc-like ridge of lower velocity emission 
became visible to the south of the other masers in the NW region.  
This region is not well-characterized by point-like emission, but is 
more diffuse and seems to be an extended ridge of maser emission.  
By the fourth epoch, the entire NW feature had grown to roughly 40 mas (12.8 AU) in extent.

We compared the positions of the reference feature, which is 
present in all epochs in the SE clump (B-REF in Table 4), 
with a maser in the NW clump that persisted through all four epochs (B-5 in Table 4).  
The separation vector for the two masers 
increased over 59.81262 days 
from 499.59 mas (159.87 AU) to 508.25 mas (162.64 AU).  A linear 
regression fit to the magnitude of the separation vector as a 
function of time yields an expansion velocity of  78.2 \kms 
with a formal error of $\pm$ 3.9 \kms. The radial velocity difference 
between the two masers of 18.5 \kms suggests an inclination of the outflow 
of about 13$^{\circ}$ from the plane of the sky.  

To investigate the proper motions of the masers associated with IRAS~4BW, 
we identified all components that were present in all four epochs of observation.  
Three components were identified in the NW and two were identified in the SE.  
We averaged the mean 
positions of the NW and SE components and averaged these two positions to 
estimate an average geometric fiducial position for each epoch. This 
position was then subtracted from each maser component.  These referenced 
positions are presented along with the fluxes and original positions in 
Table 4.  Finally, we fit straight lines to the positions 
of the components as a function of time.  The resulting tangential 
velocities and position angles from these fits are also presented in 
Table 4 and are very similar for all five maser components. 
Their motions are very uniform over the four epochs with average values of 
43 $\pm$ 2 \kms at a position angle of $-$38$^{\circ}$ $\pm$ 2$^{\circ}$. 
Figure 4 shows all of the maser components as observed in 
Epoch 1 and their corresponding proper motion vectors.  Again, an overall 
expansion is apparent along the axis connecting the two regions.  

\section{Discussion}

\subsection{IRAS~4A\label{discussion=4a}}

Since the water masers toward IRAS~4A have always been found coincident 
with the position of A2 (when they were detected), we consider 
three alternatives for the origin of the water maser emission seen using the 
VLBA:  1) the water masers are in an outflow which is the close-in extension 
of the large-scale molecular outflow, driven by IRAS~4A2; 2) the water masers 
are excited by an interaction of the close-in flow and the accretion disk 
itself and 3) the water maser outflow is driven by a companion to IRAS~4A2, 
as yet unresolved.  We consider each of these alternatives in turn.

\subsubsection{Alignment with the large scale outflow}

The alignment of the water maser outflow is curious in that it does not 
have the same position angle as any of the larger-scale outflows emanating from the 
two sources in IRAS~4A.  The axis of the large-scale outflow defined by the 
SiO, CS, CO, and H$_2$ emission \citep{Choi:2006,Choi:2005, BLAKE:1995}, 
and driven by A2, may be drifting with a rate of 0.011 degrees/yr 
(counterclockwise as seen on the sky) and a zero-point angle of around $-$166$^{\circ}$ 
(i.e., presumably the jet-injection angle), depending upon the 
fit of the large-scale emission (see Figure 3a of Choi et al. 2006). 
For the large-scale flow, the blue-shifted emission is generally in the 
south to southwest and the red-shifted emission is generally in the north and northeast.  
The maser outflow, however, on a scale of ~200 mas, has a position 
angle of $-$50$^{\circ}$, and has the more highly red-shifted emission to the southeast.
Thus, if the masers were 
marking the close-in extension to the flow, the position angle of the 
flow would have to change from $+$130$^{\circ}$ to $-$165$^{\circ}$, in a counterclockwise 
direction, from 40 to 300 AU from the star.  We consider that such a shift in 
jet direction is very unlikely, if not unphysical.

\subsubsection{Interaction with the A2 accretion disk}

The jet direction near the base of the flow, as indicated by \citet{Choi:2006}, 
and also by the CS J$=$7$\rightarrow$6 emission
\citep{BLAKE:1995} appears to be about 0 degrees (i.e. north-south).  
If the masers were excited by an interaction of the outflow with the 
A2 accretion disk itself, then the opening half-angle of the jet at its 
base would have to encompass the direction of the maser axis. Thus the 
opening half-angle of the jet would be $\sim50\deg$.  Such a large opening
angle would seem inconsistent with the Class 0 phase and the high degree of collimation seen
in the larger scale outflow.

\subsubsection{An unresolved companion to IRAS~4A2}

Perhaps the simplest explanation for the direction of the axis of the 
water maser emission is that it is not related directly at all to the 
large-scale molecular outflow, but that the masers are located in an 
independent outflow from another young stellar object.  A large 
fraction of the water masers around low-luminosity YSOs, which have been 
observed and mapped with very high angular resolution, show bipolar flows 
and structural evidence for bow shocks as the jet impinges on and sweeps up 
ambient molecular material. Although the structural evidence is sparse 
(because of the small number of maser spots) for IRAS~4A, the bipolar 
velocity separation strongly suggests that the masers are tracing a jet 
outflow as, for example, in IRAS~05413-0104 \citep{CLAUSSEN:1998}.  It is 
clear from VLA observations (e.g., Furuya et al. 2003; this paper) that the 
water masers toward IRAS~4A are always found toward A2 rather than A1, 
so A1 cannot drive the maser outflow.  Therefore we suggest, that, 
in this scenario, there must be a companion to A2 that drives 
the water masers.  Since the masers are clearly associated with the 
position of A2, and not A1, it is clear that the maser outflow, in this 
scenario, is {\bf not} associated with A1.  Indeed,
the presence of a companion to A2 separated by 30-80 AU has been proposed
by \citet{Choi:2006} to explain the drifting of the 4A2 outflow axis.  A
companion within 100 mas (32 AU) of A2 would be consistent with it not
being resolved in $\lambda$=3.6 cm continuum observations with the VLA
\citep{REIPURTH:2000}.

\subsection{IRAS~4B\label{discussion=4b}}

While the masers trace a jet similar in velocity to the HCN outflow, 
the position angle of $-$29$^{\circ}$ is significantly different from the north-south 
orientation of the outflow.  \citet{Choi:2001} suggests a dynamical lifetime 
for the molecular outflow of $\sim$200 years (corrected to a distance of 320 pc).  
This is much longer than the dynamical lifetime of the maser emission 
which is on the order of a year.  We conclude that the jet must be precessing 
in such a way as to move the projected outflow axis from a position angle 
of 0$^{\circ}$ to $-$29$^{\circ}$ in several hundred years.  In this picture, the 
jet would be impacting the walls of the outflow cavity as it precessed 
``clockwise" in the plane of the sky.  The NW maser group would represent 
the western wall of the pre-existing cavity and the SE maser group the 
eastern wall.  This would explain the linear, arc-like structure of the 
maser groups and the appearance of new masers at the southern extent of the NW group.  
This model can only be reconciled with observations of dense outflowing gas traced 
in H$_2$CO by \citet{DIFRANCESCO:2001} if there were a second outflow from an 
unresolved companion to IRAS~4BW.
If our picture is correct, one would expect future VLBI observations to 
reveal new masers south (north) of the NW (SE) group.  One would also 
expect that on a larger scale, tracers of the jet would display a 
characteristic S-shape, bending back to a north-south orientation.  
Indeed, slow precessing jets 
have been proposed for a number of YSO outflows \citep{Matthews:2006,Hodapp:2005,
Terquem:1999} that suggest the presence of a nearby companion \citep{Terquem:1998}.

The proper motion vectors for the five persistent maser spots relative to a 
fiducial point (Figure 4) suggests a larger position angle than that of the outflow axis 
defined by the relative motion of SE and NW masers ($-$38$^{\circ}$ vs. $-$29$^{\circ}$, see Table 4).  
This is reminiscent of the proper motions seen in the triple radio continuum source
in Serpens \citep{Curiel:1993}, where the outer components of the radio jet have the
same tangential velocity but are moving in slightly different directions. Although the
maser emission in IRAS~4B traces the jet at scales 40 times smaller than that traced by
the radio continuum in the Serpens jet, this difference in relative motions may be 
suggestive of similar physical processes in the two jets.

The larger position angle defined by the proper motion vectors in IRAS~4B suggest that, 
in addition to the tangential velocity imparted to the 
masing gas by the outflow jet at a position angle of $-$29$^{\circ}$, there is an 
additional component of about 6 \kms perpendicular to the outflow axis.  
It is possible that this motion is related to the expansion of the outflow cavity into the ambient cloud.  

\subsection{Maser Geometry}

We estimate that the maser outflows in both IRAS~4A and 4B are nearly in the plane of the sky 
(inclination of only 2$^{\circ}$ for IRAS~4A and about 13$^{\circ}$ for IRAS~4B).  In
Claussen et al. (1998) we estimated that the inclination of the maser outflow for IRAS~05413-0104
was only 4$^{\circ}$.  Models of maser emission (e.g. Hollenbach 1997) suggest that the
masers trace shocks produced by the interaction of the jet and ambient material along the
jet.  Shocks that propagate close to the plane of the sky provide much longer maser coherence
lengths than those that propagate at larger inclination angles.  Thus outflows that have 
inclinations close to the plane of the sky provide a more favorable maser geometry, perhaps
explaining why some well-known outflows from low-mass YSOs do not show maser emission.

\section{Conclusions}

We have observed the water masers associated with IRAS~4A and 4B at VLBI resolutions in four epochs
over three months.  We have determined that the masers are related to the jets emanating from these 
YSOs due to their spatio-kinematic distribution.  In both sources the masers are found associated with known single components of multiple systems, with total separation velocities between 53 and 78 km s$^{-1}$.

This is further confirmed by the large proper motions measured
for both sources, which clearly rules out rotation due to the mass constraints placed on the
central objects by other observations.  The water masers of IRAS~4B form arc-like structures
roughly 10 AU in length and less than 0.6 AU in thickness.  The structure of these structures changes
rapidly with time, with many new maser components appearing and disappearing in just one month.  The orientation of the structures in the plane of the sky does not agree with larger scale outflow angle, which we attribute to a possible unseen very close companion.
Future observations will have to sample the source more frequently than once every three weeks, with
once every 3-5 days probably providing the best results.

\acknowledgments

KBM wishes to thank the AAS Council and AAS Executive Officer, 
Dr. Robert Milkey for allowing him to continue research work in 
his current position.  KBM thanks TEK for her continued support.
And special thanks to T. Beasley for exposing KBM to the rigors of
Outback observing and a grueling exercise regimen, during which
time a portion of this paper was written.  Finally, KBM thanks
the NRAO for hosting him during a research leave from the AAS, during
which time the bulk of this paper was completed.

\appendix
\section{Table of Maser Components for IRAS~4B}
Maser component fits were averaged using a flux-squared weighting scheme, that
emphasizes the strongest portion of the maser emission and produces a reliable location for
the maser.  As some masers are present for only three channels, a full Gaussian fit to the
maser as a function of velocity was deemed impractical, although it could have been used for
the masers that existed over many channels.  We opted to use the same position-determining
scheme for all features.  These averaged component fits and their corresponding
errors are given in Table 5 for IRAS~4B.  Note that the uncertainty in position for the reference
channel is simply the error on the mean position values determined for the spot and is
indicative of the reliability of the Gaussian fitting routines in the presence of noise
in the images.

%
%
\clearpage
%
\begin{table}
\caption{IRAS~4A and IRAS~4B Properties}
\begin{center}\scriptsize 
\begin{tabular}{lcc} Property & Value & Reference \\ 
\tableline \tableline  IRAS4A  &  \\ 
A1 2.7 mm Position (J2000.0)  & 3$^h$ 29$^m$ 10$^s$.510  +31$^o$ 13\arcmin 31\arcsec.01 & Looney, Mundy and Welch 2000 \\ 
A1 1.3 cm Position (J2000.0)  & 3$^h$ 29$^m$ 10$^s$.527  +31$^o$ 13\arcmin 31\arcsec.06 & This study \\
A1 3.6 cm Position (J2000.0)  & 3$^h$ 29$^m$ 10$^s$.529  +31$^o$ 13\arcmin 31\arcsec.05 & Reipurth \etal 2002 \\
A2 2.7 mm Position (J2000.0)  & 3$^h$ 29$^m$ 10$^s$.413  +31$^o$ 13\arcmin 32\arcsec.20 & Looney, Mundy and Welch 2000 \\ 
A2 3.6 cm Position (J2000.0)  & 3$^h$ 29$^m$ 10$^s$.421  +31$^o$ 13\arcmin 32\arcsec.21 & Reipurth \etal 2002 \\  
A2 H$_2$O NW Position (J2000.0)  & 3$^h$ 29$^m$ 10$^s$.4101  +31$^o$ 13\arcmin 32\arcsec.235 & VLA-12Aug98; This study\\
A2 H$_2$O SE Position (J2000.0)  & 3$^h$ 29$^m$ 10$^s$.4226  +31$^o$ 13\arcmin 32\arcsec.142 & VLA-12Aug98; This study\\
A1 Peak Flux (2.7 mm)  & 107 mJy &  (2\arcsec) Looney, Mundy, \& Welch 2000 \\
A2 Peak Flux (2.7 mm)  & 23 mJy & (2\arcsec)   Looney, Mundy, \& Welch 2000  \\
A1 Peak Flux (3.6 cm)  & 0.32 mJy &  Reipurth \etal 2002 \\
A2 Peak Flux (3.6 cm)  & 0.11 mJy &  Reipurth \etal 2002 \\
Bolometric Luminosity (L$_\odot$) &     14     & Andr\'e, Ward-Thompson, \& Barsony 2000\\
Bolometric Temperature (K) &     34      &  Andr\'e, Ward-Thompson, \& Barsony 2000\\
\tableline
 IRAS4B  &  \\
2.7 mm Position (J2000.0)  & 3$^h$ 29$^m$ 11$^s$.988  +31$^o$ 13\arcmin 08\arcsec.10 & Looney, Mundy, \& Welch 2000 \\  
1.3 cm Position (J2000.0)  & 3$^h$ 29$^m$ 12$^s$.001  +31$^o$ 13\arcmin 08\arcsec.16 & This study \\  
3.6 cm Position (J2000.0)  & 3$^h$ 29$^m$ 12$^s$.003  +31$^o$ 13\arcmin 08\arcsec.14 & Reipurth \etal 2002\\  
B H$_2$O NW Position (J2000.0)  & 3$^h$ 29$^m$ 11$^s$.9923  +31$^o$ 13\arcmin 08\arcsec.363 & VLA-12Aug98; This study\\
B H$_2$O SE Position (J2000.0)  & 3$^h$ 29$^m$ 12$^s$.0104  +31$^o$ 13\arcmin 07\arcsec.955 & VLA-12Aug98; This study\\  
Peak Flux (2.7 mm)  & 58 mJy & (2\arcsec) Looney, Mundy, \& Welch 2000 \\
Peak Flux (3.6 cm)  & 0.33 mJy &  Reipurth \etal 2002 \\
Bolometric Luminosity (L$_\odot$) &     14     & Andr\'e, Ward-Thompson, \& Barsony 2000\\
 Bolometric Temperature (K) &     36      &  Andr\'e, Ward-Thompson, \& Barsony 2000\\
\tableline
\end{tabular}
\end{center}
\end{table}

%
\clearpage
%
%
\begin{deluxetable}{lllrr}
%
%
\tablewidth{0pt}
\tablecaption{Dates and times of Observations \label{tab2}}
\tablehead{
\colhead{Epoch} & \colhead{Date} & \colhead{Mid-time} & \colhead{Julian Date} 
& \colhead{$\Delta$Time [d]} \\
\colhead{} & \colhead{of 1998} & \colhead{UT} & \colhead{-2451000.0}
} 
\startdata
 I   & September 15 & 11:01:23   &  71.95928 &  0.00000  \\
 II  & October 4    & 10:01:30   &  90.91771 & 18.95843 \\
 III & October 27   & 08:01:30   & 113.83437 & 41.87509   \\
 IV  & November 14  & 06:31:30   & 131.77187 & 59.81259  \\
\enddata
\end{deluxetable}

%
\clearpage
%
%
\begin{deluxetable}{lcrr}
%
%
\tablewidth{0pt}
\tablecaption{Imaging parameters for each epoch \label{tab3}}
\tablehead{
\colhead{Epoch} & \colhead{Beam Dimensions} & \colhead{Beam Position Angle} & \colhead{RMS} \\
\colhead{} & \colhead{[$\mu$arcsec.]} & \colhead{[$^{\circ}$ E of N]} & \colhead{[mJy/Beam]} \\
} 
\startdata
 4A-I    & 1260 $\times$ 910 &  11.8 & 10.4 \\
 4A-II   &  830 $\times$ 440 & -21.2 &  6.8 \\
 4A-III  &  810 $\times$ 400 & -19.1 & 11.7 \\
 4A-IV   &  650 $\times$ 370 &  -8.2 &  9.6 \\
& & & \\
4B-I     & 650 $\times$ 350 & -14.4  & 13.9 \\
4B-II    & 620 $\times$ 330 & -12.4  &  7.6 \\
4B-III   & 590 $\times$ 330 & -10.8  &  9.9 \\
4B-IV    & 720 $\times$ 420 &  -5.8  &  6.0 \\
\enddata
\end{deluxetable}

%
%
\begin{deluxetable}{lrrrrrr}
%
%
\tablewidth{0pt}
\tablecaption{Proper Motions for Masers Observed in All Four Epochs \label{tab4}}
\tablehead{

\colhead{Epoch ID} & \colhead{Velocity} & \colhead{Peak Flux} & \colhead{X-Offset} & \colhead{X$_{ref}$}& \colhead{Y-Offset} & \colhead{Y$_{ref}$} \\
\colhead{} & \colhead{(km sec$^{-1}$)} & \colhead{(Jy)} & \colhead{(mas)} & \colhead{(mas)} & \colhead{(mas)} & \colhead{(mas)} 

}

\startdata
{\bf IRAS 4A} & & & & & & \\
I-A-REF\tablenotemark{a}   & 10.40 & 3.29  &  0.083 & \nodata &  0.073  & \nodata \\
II-A-REF  & 10.55 & 4.14  & -0.063 & \nodata & -0.185  & \nodata \\
III-A-REF & 10.68 & 7.97  &  0.007 & \nodata & -0.030  & \nodata \\
IV-A-REF  & 10.72 & 14.9  & -0.005 & \nodata &  0.002  & \nodata \\
\\
I-A-1   & 7.61   & 0.969  & -167.543 & -167.626 &  187.223  & 187.150 \\
II-A-1  & 7.65   & 0.211  & -169.013 & -168.950 &  187.985  & 188.170 \\
III-A-1 & 7.71   & 0.208  & -170.631 & -170.638 &  189.606  & 189.636 \\
IV-A-1  & 7.85   & 0.123  & -172.098 & -172.093 &  190.924  & 190.921 \\
v$_{exp}$ & 53.02   \\
PA(Deg) &-49.82\arcdeg   \\
\\
I-A-2   & 10.34   & 1.90  & -1.859 & -1.942 & -0.086  &-0.160 \\
II-A-2  & 10.36   & 2.26  & -1.906 & -1.844 & -0.118  & 0.067 \\
III-A-2 & 10.29   & 1.74  & -1.844 & -1.851 &  0.695  & 0.725 \\
IV-A-2  & 10.35   & 1.27  & -1.938 & -1.933 &  1.113  & 1.108 \\
\\
{\bf IRAS 4B} & & & & & & \\
I-B-REF\tablenotemark{b}   & -0.750   & 36.8 & -0.020 & 122.668 & 0.0782 & -227.845 \\
II-B-REF  & -1.161   & 20.4 & -0.001 & 123.598 &-0.041  & -228.993 \\
III-B-REF & -0.224   & 56.6 & -0.000 & 124.475 &-0.013  & -229.284 \\
IV-B-REF  & -0.273   & 34.3 & -0.002 & 125.498 & 0.010  & -230.661 \\
\\
I-B-1     & 0.380    & 0.368& -1.736 & 120.932  & 10.016 & -217.830  \\
II-B-1    & 0.302    & 2.14 & -1.677 & 121.921  & 9.384  & -219.549  \\
III-B-1   & 0.485    & 0.256& -1.626 & 122.849  & 8.726  & -220.559  \\
IV-B-1    &-0.158    & 0.09 & -1.732 & 123.767  & 8.912  & -221.749  \\
v$_{exp}$ & 45.40   \\
PA(Deg)   &144.12\arcdeg \\
\\
I-B-2     & 2.81     & 0.829& -9.429 & 113.239 & 33.556 & -194.290  \\
II-B-2    & 3.05     & 0.134& -9.531 & 114.067 & 33.844 & -195.089  \\
III-B-2   & 3.25     & 1.24 & -9.074 & 115.402 & 32.605 & -196.679  \\
IV-B-2    & 3.39     & 0.218& -9.417 & 116.082 & 32.975 & -197.686  \\
v$_{exp}$ & 41.03  \\
PA(Deg)   &140.07\arcdeg \\
\\
I-B-3     & 17.46 & 0.634 & -239.832 & -117.165 & 432.737 & 204.892  \\
II-B-3    & 17.07 & 2.80  & -241.622 & -118.024 & 434.953 & 206.020  \\
III-B-3   & 17.27 & 0.521 & -243.675 & -119.200 & 436.448 & 207.164  \\
IV-B-3    & 17.13 & 0.318 & -245.662 & -120.163 & 439.063 & 208.402  \\
v$_{exp}$ & 41.90  \\
PA(Deg)  &-40.50\arcdeg  \\
\\
I-B-4     & 17.74 & 2.31  & -239.610 & -116.943 & 430.618 & 202.772  \\
II-B-4    & 17.66 & 5.90  & -241.452 & -117.853 & 432.924 & 203.991  \\
III-B-4   & 17.60 & 1.50  & -243.359 & -118.884 & 434.451 & 205.167  \\
IV-B-4    & 17.43 & 0.862 & -245.111 & -119.612 & 437.038 & 206.377  \\
v$_{exp}$ & 41.47 \\
PA(Deg) &-36.51\arcdeg  \\
\\
I-B-5   & 17.95 & 0.643  & -239.816 & -117.148 & 438.361 & 210.516  \\
II-B-5  & 17.99 & 0.957  & -241.704 & -118.106 & 440.878 & 211.947  \\
III-B-5 & 17.84 & 1.24   & -243.767 & -119.292 & 442.811 & 213.527  \\
IV-B-5  & 17.22 & 1.55   & -245.495 & -119.997 & 445.035 & 214.374  \\
v$_{exp}$ & 46.15   \\
PA(Deg) &-36.44\arcdeg   \\ 
\enddata
\tablenotetext{a}{Strongest maser in IRAS~4A used as both a position and velocity reference.}
\tablenotetext{b}{Strongest maser in IRAS~4B used as a position reference only. Proper motions 
are measured relative to a fiducial point for each epoch given as X$_{ref}$ and Y$_{ref}$
for this maser (see text).}
\end{deluxetable}

\clearpage
%
%
\begin{deluxetable}{llrrrrrr}
%
%
\tablewidth{0pt}
\tablecaption{Fitted Gaussian Component in IRAS 4B \label{tab5}}
\tablehead{
\colhead{Epoch ID} &  \colhead{Other} & \colhead{Velocity} & \colhead{Peak Flux} & \colhead{X-Offset} & \colhead{Y-Offset} & \colhead{$\sigma_{x}$} 
& \colhead{$\sigma_y$} \\
\colhead{} &  \colhead{} & \colhead{(\kms)} & \colhead{(Jy)} & \colhead{(mas)} & \colhead{(mas)} & \colhead{($\mu$as)} & \colhead{($\mu$as)} \\
} 
\startdata
I-SE-1   &B-2&  2.81 &  0.83  & -9.428      & 33.555  & 4.4   & 8.0  \\
I-SE-2   &    &  2.30 &  8.24 & -7.362      & 27.891  & 0.1   & 1.1  \\
I-SE-3   &    &  2.21 &  4.97 & -5.722      & 21.858   & 1.1   & 1.3  \\
I-SE-4   &    &  2.27 &  0.76 & -5.966      & 22.825   & 5.5   & 5.3  \\
I-SE-5   &    &  2.10 &  0.44 & -6.433      & 23.448   & 7.8   & 12.9 \\
I-SE-6   &    &  1.94 &  0.76 & -5.327      & 20.908   & 3.8   & 5.5  \\
I-SE-7  &    &  1.88 &  1.76  & -5.883      & 22.016   & 1.2   & 2.5  \\
I-SE-8  &    &  1.74 &  0.58  & -5.247      & 20.739   & 6.1   & 11.8 \\
I-SE-9  &    &  1.33 &  0.37  & -4.515      & 18.473   & 7.6   & 14.6 \\
I-SE-10  &    &  1.29 &  0.38 & -3.104      & 15.207   & 8.1   & 14.7 \\
I-SE-11  &B-1 &  0.38 &  0.37 & -1.735      & 10.010    & 6.6   & 8.8  \\
I-SE-12  &    &  0.23 &  0.62 & -1.442      & 8.355    & 3.7   & 2.9  \\
I-SE-13  &    &  0.16 &  5.84 & -1.571      &  9.156   & 0.3   & 1.2  \\
I-SE-14  &    &  0.10 &  0.41 & -1.281      & 6.440    & 6.7   & 12.1 \\
I-SE-15  &    & -0.01 &  0.48 & -1.392      & 7.743    & 7.2   & 21.2 \\
I-SE-16 &B-REF& -0.75& 36.76  & -0.020360   & 0.07815  & 0.007 & 0.04 \\
I-SE-17   &    & -1.57 &  0.20 & -49.551     & 88.150   & 14.0  & 25.6 \\
I-SE-18   &    & -2.68 &  0.38 & -48.231     & 84.987   & 7.42  & 13.1 \\
I-NW-1    &B-5 & 17.95 &  0.64 & -239.816    & 438.361  & 4.8   & 9.4  \\
I-NW-2    &B-4 & 17.74 &  2.31 & -239.610    & 430.618  & 1.4   & 3.5  \\
I-NW-3    &    & 17.71 &  0.39 & -239.404    & 429.748  & 6.2   & 4.5  \\
I-NW-4    &    & 17.70 &  0.27 & -239.354    & 427.401  & 8.8   & 23.2 \\
I-NW-5    &B-3 & 17.46 &  0.63 & -239.832    & 432.737  & 5.1   & 2.2  \\
I-NW-6    &    & 17.34 &  8.64 & -239.793    & 431.960  & 0.1   & 1.0  \\
I-NW-7    &    & 17.31 &  0.52 & -239.674    & 431.152  & 4.5   & 2.4  \\
I-NW-8    &    & 17.13 &  2.13 & -239.178    & 420.144  & 1.6   & 5.1  \\
I-NW-9    &    & 16.93 &  3.15 & -239.996    & 425.710  & 1.3   & 3.3  \\
I-NW-10   &    & 16.88 &  0.72 & -240.000    & 424.172  & 5.2   & 22.2 \\
I-NW-11   &    & 16.71 &  0.46 & -239.842    & 421.856  & 7.3   & 24.4 \\
I-NW-12   &    & 16.28 &  0.23 & -239.074    & 413.088  & 14.8  & 31.3 \\ 
I-NW-13   &    & 16.11 &  0.12 & -239.817    & 415.178  & 28.7  & 54.5 \\ 
         &    &       &        &             &          &       &       \\           
II-SE-1  &    &  3.08 &  0.13 & -9.929   &  34.227 & 18    & 28 \\
II-SE-2  &B-2 &  3.05 &  0.13 & -9.532   &  33.844 & 18    & 23 \\
II-SE-3  &    &  2.53 &  0.32 & -6.009   &  22.195 &  6    & 10 \\
II-SE-4  &    &  2.38 &  1.09 & -7.803   &  28.253 &  2    &  3 \\
II-SE-5  &    &  2.26 &  0.46 & -6.057   &  22.259 &  5    &  7 \\
II-SE-6  &    &  2.24 &  0.23 & -6.545   &  23.227 & 10    & 17 \\
II-SE-7  &    &  1.63 &  0.18 & -2.539   &  15.374 & 11    & 21 \\
II-SE-8  &    &  1.60 &  0.48 & -4.945   &  19.856 &  5    &  9 \\
I-SE-9  &    &  1.50 &  0.64 & -4.899   &  19.603 &  3    &  1 \\
II-SE-10  &    &  1.48 &  1.23 & -4.751   &  18.793 &  2    &  5 \\
II-SE-11 &    &  1.22 &  0.25 & -3.393   &  15.304 & 10    & 17 \\
II-SE-12 &    &  1.13 &  0.14 & -2.706   &  13.848 & 19    & 30 \\
II-SE-13 &    &  1.10 &  0.17 & -2.724   &  13.181 & 10    & 22 \\
II-SE-14 &    &  0.84 &  0.53 & -2.317   &  11.923 &  4    & 10 \\
II-SE-15 &    &  0.43 &  2.44 & -1.758   &   9.760 &  1    &  1 \\
II-SE-16 &    &  0.43 &  1.05 & -1.558   &   8.746 &  1    &  1 \\
II-SE-17 &B-1 &  0.30 &  2.14 & -1.677   &   9.383 &  1    &  2 \\
II-SE-18 &    &  0.21 &  1.03 & -1.298   &   6.382 &  2    &  4 \\
II-SE-19 &    &  0.11 &  0.29 & -0.580   &   2.781 &  6    & 13 \\
II-SE-20 &    &  0.05 &  0.70 & -1.401   &   7.519 &  2    &  5 \\
II-SE-21 &    & -0.97 & 12.80 & -0.132670 &   0.679741 &  0.002   &  0.005 \\
II-SE-22 &B-REF& -1.16 & 20.41 & -0.000980 &  -0.041125 &  0.004  &  0.007 \\
II-SE-23 &    & -1.52 &  2.11 &  0.244   &  -1.140 & 1   &  1 \\
II-SE-24  &    & -1.90 &  0.32 & -46.0974 &  77.935 &  6    & 11 \\
II-SE-25 &    & -2.04 &  0.24 &  1.360   &  -6.688 &  6    & 11 \\
II-SE-26  &    & -2.37 &  0.29 & -47.5718 &  82.438 &  7    & 17 \\
II-NW-1   &B-5 & 17.99 &  0.96 & -241.704 & 440.878 &  2    &  5 \\
II-NW-2   &    & 17.99 &  0.86 & -241.869 & 449.662 &  2    &  4 \\
II-NW-3   &B-4 & 17.66 &  5.90 & -241.452 & 432.924 &  1    &  1 \\
II-NW-4   &    & 17.61 &  1.48 & -241.606 & 433.736 &  2    & 10 \\
II-NW-5  &    & 17.10 &  0.34 & -241.901 & 429.900 &  9    & 38 \\
II-NW-6  &    & 17.09 &  0.56 & -241.571 & 427.050 &  5    &  5 \\
II-NW-7   &B-3 & 17.07 &  2.80 & -241.622 & 434.953 &  1    &  1 \\
II-NW-8   &    & 17.04 &  3.58 & -241.631 & 428.024 &  1    &  2 \\
II-NW-9  &    & 17.03 &  0.26 & -241.059 & 423.332 & 11    &  6 \\
II-NW-10   &    & 17.02 &  3.29 & -241.073 & 422.442 &  1    &  1 \\
II-NW-11  &    & 16.97 &  0.37 & -241.235 & 421.489 &  6    &  7 \\
II-NW-12  &    & 16.95 &  0.26 & -241.421 & 424.037 &  9    & 28 \\
II-NW-13   &    & 16.90 &  0.43 & -241.797 & 429.376 &  6    & 12 \\
II-NW-14   &    & 16.87 &  0.54 & -241.683 & 426.113 &  5    & 19 \\
         &    &       &        &       &         &     &      \\           
III-SE-1  &B-2&  3.25 &  1.24 &   -9.074  &  32.605 &  2  &  2 \\ 
III-SE-2  &   &  3.19 &  0.32 &   -4.623  &  22.469 &  6  & 11 \\  
III-SE-3  &   &  1.83 &  0.18 &   -2.942  &  16.312 & 11  & 17 \\
III-SE-4  &   &  1.65 &  0.18 &   -4.944  &  18.837 &  9  & 23 \\
III-SE-5  &   &  1.38 &  0.14 &   -4.707  &  17.604 & 12  & 22 \\
III-SE-6  &   &  0.99 &  0.81 &   -2.259  &  10.991 &  3  &  5 \\
III-SE-7  &   &  0.60 &  0.18 &   -1.151  &   6.024 & 10  & 22 \\
III-SE-8  &   &  0.51 &  0.12 &   -1.388  &   7.749 & 13  & 23 \\
III-SE-9  &B-1&  0.48 &  0.26 &   -1.626  &   8.726 &  8  & 18 \\
III-SE-10 &   &  0.30 &  0.16 &   -1.118  &   5.494 &  9  & 28 \\
III-SE-11 &   &  0.09 &  0.73 &   -0.421  &   1.823 &  3  &  7 \\
III-SE-12 &B-REF& -0.22 & 56.60 &   -0.000320 &  -0.013123 & 0.001 & 0.005 \\
III-SE-13 &   & -0.26 &  2.17 &    0.134  &  -0.762 &  1  &  1 \\
III-SE-14 &   & -0.67 &  0.22 &    1.200  &  -4.693 & 11  & 21 \\
III-SE-15 &   & -0.84 &  2.92 &  -45.799  &  75.965 &  1  &  1 \\
III-SE-16 &   & -0.97 &  0.16 &    2.054  &  -8.797 &  9  & 29 \\
III-SE-17 &   & -1.01 &  0.13 &    0.504  &  -2.326 & 14  & 21 \\
III-SE-18 &   & -1.34 &  1.33 &    1.165  &  -5.231 &  1  &  3 \\
III-SE-19 &   & -1.34 &  0.23 &    0.947  &  -4.513 &  7  &  9 \\
III-SE-20 &   & -1.69 &  0.20 &  -45.966  &  76.791 & 10  & 11 \\
III-SE-21 &   & -2.72 &  0.39 &  -48.936  &  84.458 &  6  & 10 \\
III-NW-1  &   & 17.99 &  3.58 & -243.777  & 451.560 & 0.3 &  1 \\
III-NW-2  &B-5& 17.84 &  1.24 & -243.767  & 442.811 &  2  &  4 \\
III-NW-3  &   & 17.64 &  0.52 & -243.826  & 442.337 &  5  & 10 \\
III-NW-4  &B-4& 17.60 &  1.50 & -243.359  & 434.451 &  1  &  3 \\
III-NW-5  &   & 17.34 &  0.99 & -244.031  & 441.137 &  2  &  4 \\
III-NW-6  &   & 17.28 &  1.26 & -243.607  & 435.704 &  2  &  4 \\
III-NW-7  &B-3& 17.27 &  0.52 & -243.675  & 436.448 &  4  &  6 \\
III-NW-8  &   & 17.14 &  0.31 & -243.397  & 430.521 &  8  & 23 \\
III-NW-9  &   & 17.02 &  1.00 & -243.439  & 429.509 &  3  &  5 \\
III-NW-10 &   & 16.94 &  0.60 & -243.038  & 424.096 &  3  &  6 \\
III-NW-11 &   & 16.86 &  0.14 & -243.471  & 427.598 & 13  & 33 \\
III-NW-12 &   & 16.83 &  0.30 & -243.662  & 431.119 & 10  & 24 \\
III-NW-13 &   & 16.49 &  0.39 & -244.642  & 437.123 &  6  & 13 \\
III-NW-14 &   & 16.20 &  0.10 & -244.699  & 434.707 & 25  & 84 \\ 
         &          &        &           &         &           &      \\           
IV-SE-1  &B-2&  3.39 &  0.22 &   -9.417 &  32.975 & 7  & 12 \\
IV-SE-2  &   &  2.30 &  0.48 &   -0.527 &  10.169 & 4  &  7 \\
IV-SE-3  &   &  1.18 &  0.08 &    1.014 &   2.647 & 22 & 36 \\
IV-SE-4  &B-1& -0.16 &  0.09 &   -1.732 &   8.912 & 17 & 25 \\
IV-SE-5  &B-REF& -0.27 & 34.28 &   -0.001790 & 0.009699  & 0.001 & 0.004  \\
IV-SE-6  &   & -1.46 &  0.09 &    2.713 & -10.796 & 19 & 33 \\
IV-SE-7  &   & -1.52 &  0.31 &  -46.019 &  76.637 &  5 & 10 \\
IV-SE-8  &   & -1.65 &  0.07 &  -46.833 &  78.663 & 24 &  5 \\
IV-NW-1  &   & 19.07 &  0.19 & -245.612 & 458.674 &  7 & 13 \\
IV-NW-2  &   & 18.99 &  2.04 & -245.643 & 456.618 &  1 &  1 \\
IV-NW-3  &   & 18.80 &  0.48 & -245.576 & 458.611 &  3 & 14 \\
IV-NW-4  &   & 18.60 &  0.47 & -245.375 & 460.263 &  3 &  5 \\
IV-NW-5  &   & 18.29 &  0.30 & -245.403 & 453.824 &  4 &  8 \\
IV-NW-6  &   & 17.81 &  0.59 & -245.703 & 449.530 &  3 &  6 \\
IV-NW-7  &B-4& 17.43 &  0.86 & -245.111 & 437.038 &  2 &  6 \\
IV-NW-8  &B-5& 17.22 &  1.55 & -245.495 & 445.035 &  1 &  2 \\
IV-NW-9  &   & 17.13 &  0.92 & -245.701 & 443.219 &  2 &  4 \\
IV-NW-10 &B-3& 17.13 &  0.32 & -245.662 & 439.063 &  9 & 14 \\
IV-NW-11 &   & 17.09 &  0.35 & -245.643 & 444.135 &  6 &  8 \\
IV-NW-12 &   & 17.05 &  0.61 & -246.666 & 446.387 &  3 &  8 \\
IV-NW-13 &   & 16.27 &  0.14 & -246.060 & 440.124 & 15 & 16 \\
IV-NW-14 &   & 16.14 &  0.99 & -246.162 & 439.176 &  2 &  4 \\
IV-NW-15 &   & 14.33 &  0.52 & -247.708 & 425.411 &  2 &  7 \\
\enddata
\end{deluxetable}

\clearpage
%
%
%
%
\begin{figure}
\includegraphics[scale=0.25]{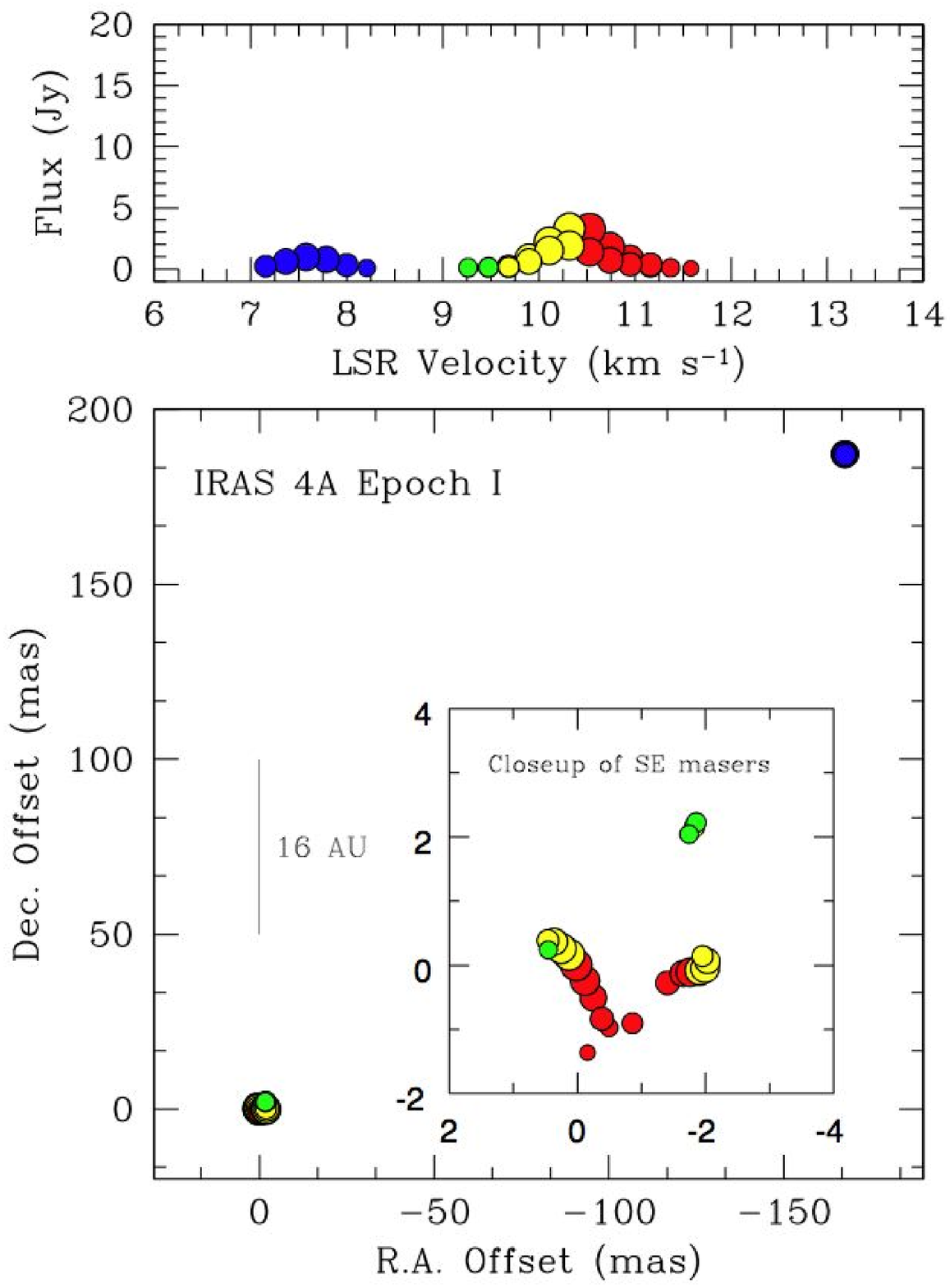}
\includegraphics[scale=0.25]{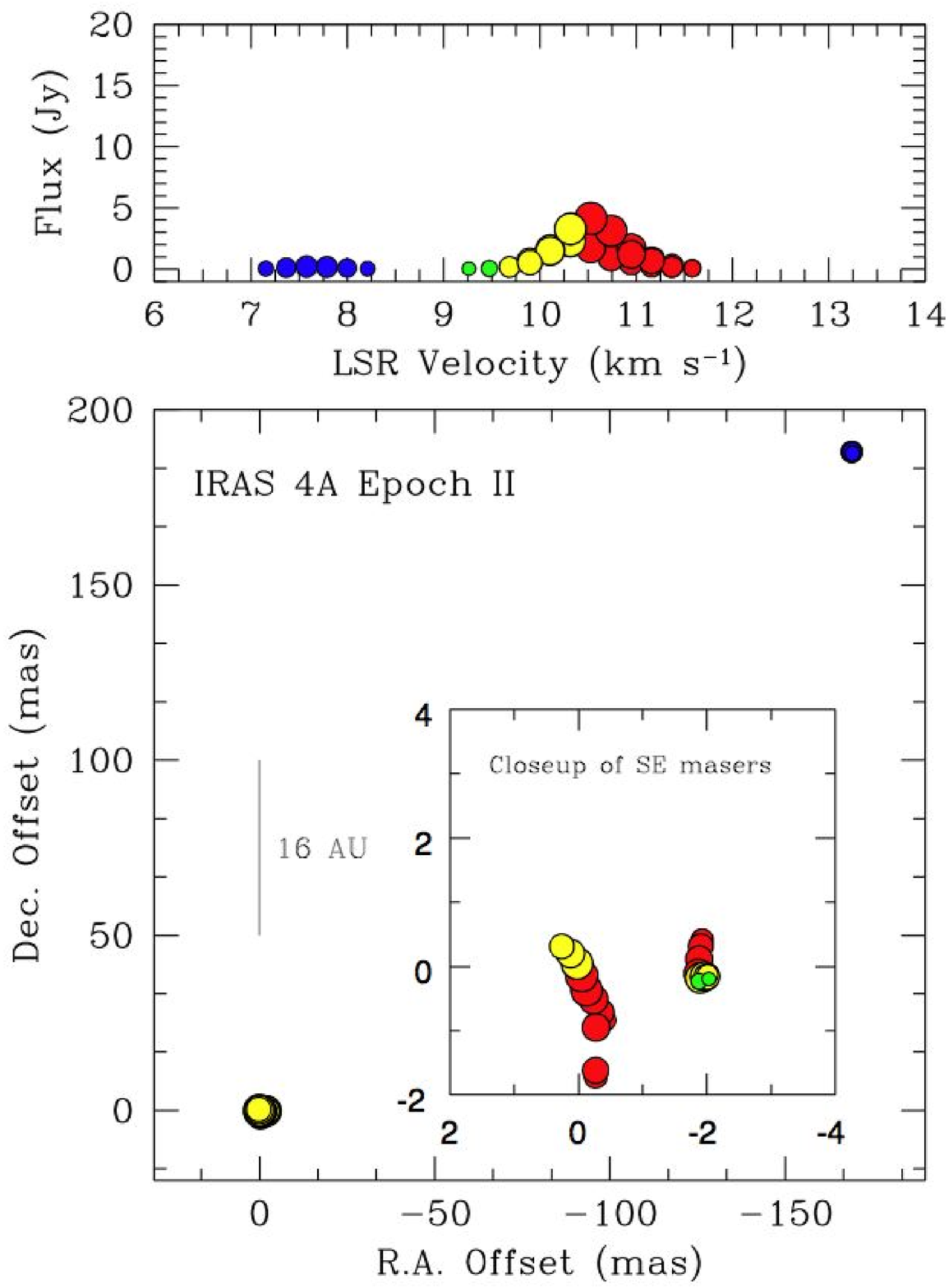}
\includegraphics[scale=0.25]{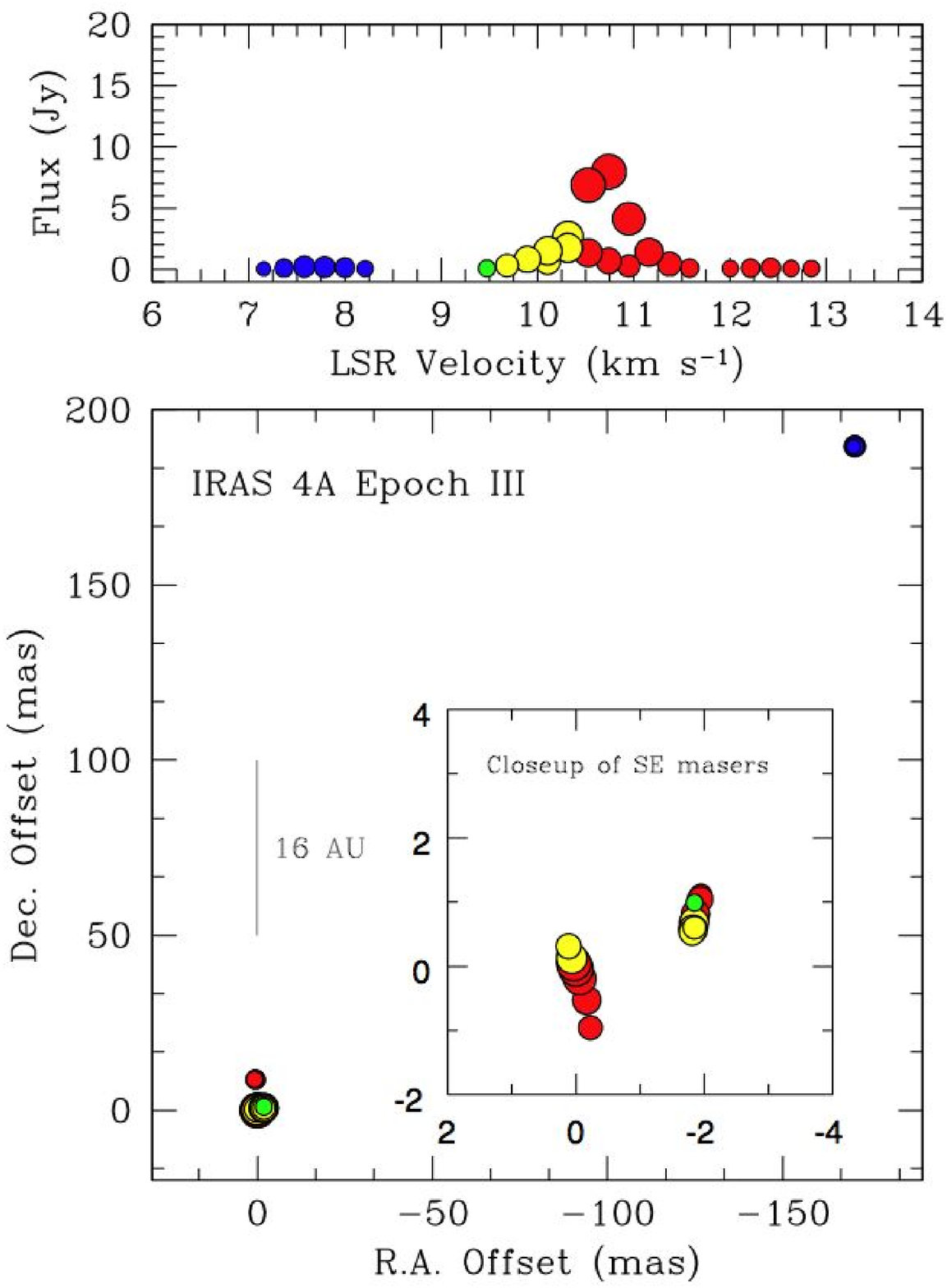}
\includegraphics[scale=0.25]{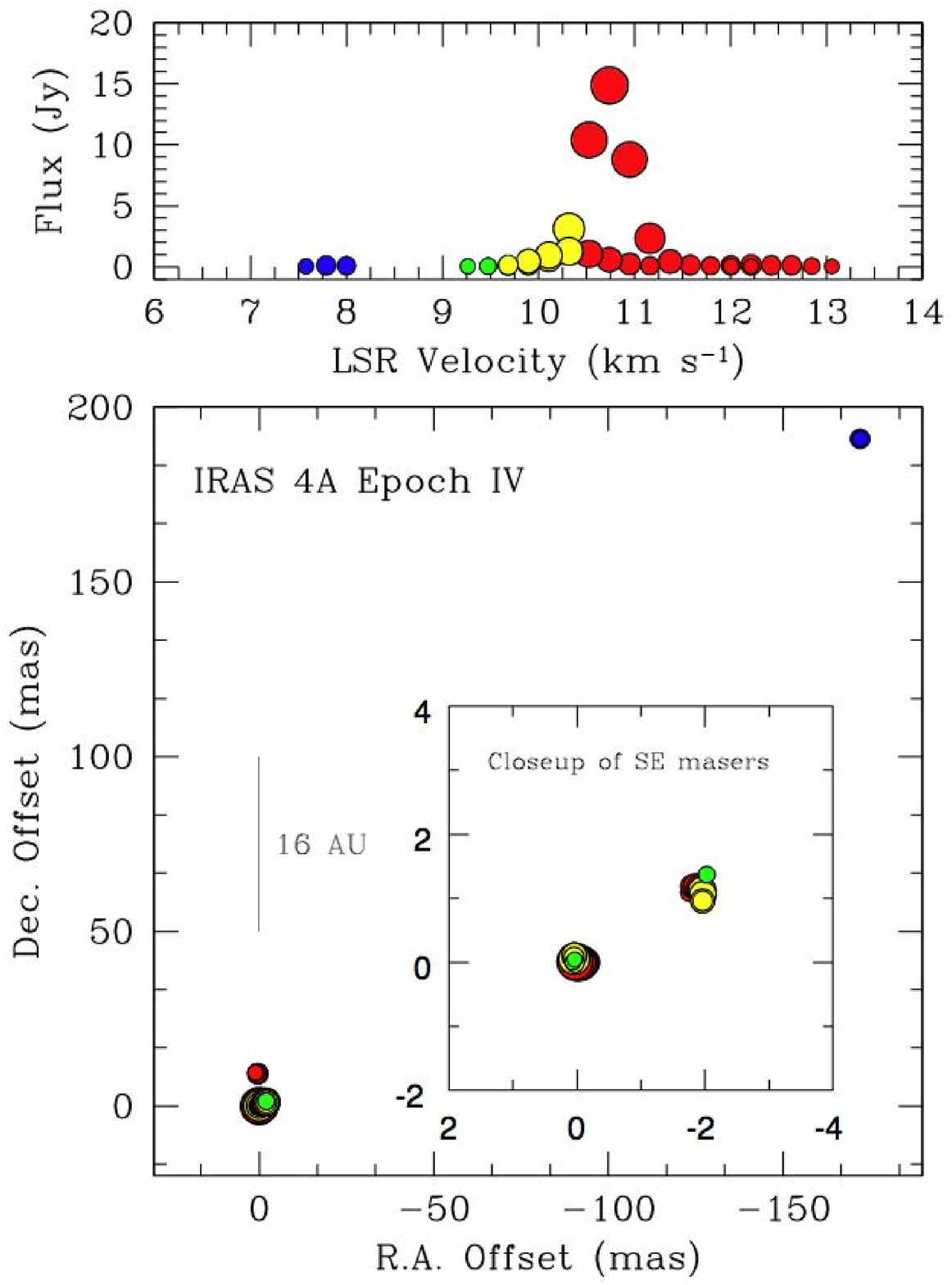}

\caption{
The spatial and kinematic distribution of the water masers associated with IRAS~4A2 for the
four epochs observed.  The top frame in each panel shows the velocity and flux of the detected 
features, the lower frame shows the
spatial distribution. The 
maser  to the NW is colored blue (at the cloud velocity) and those to the SE are coded red 
(red-shifted relative to the cloud velocity).  The size of the circles
in the figure is proportional to the flux of that maser component.  All components measured in
each channel are presented (e.g., no component averaging has been performed).  The error in the 
position of the spots (typically less than 20 $\mu$arcseconds) cannot be plotted on the large scale 
figure.  
}
\notetoeditor{This four-panel color figure should appear on a single page and
each of the four panels should be legible.}
\end{figure}
\clearpage
\begin{figure}
\includegraphics[scale=0.43]{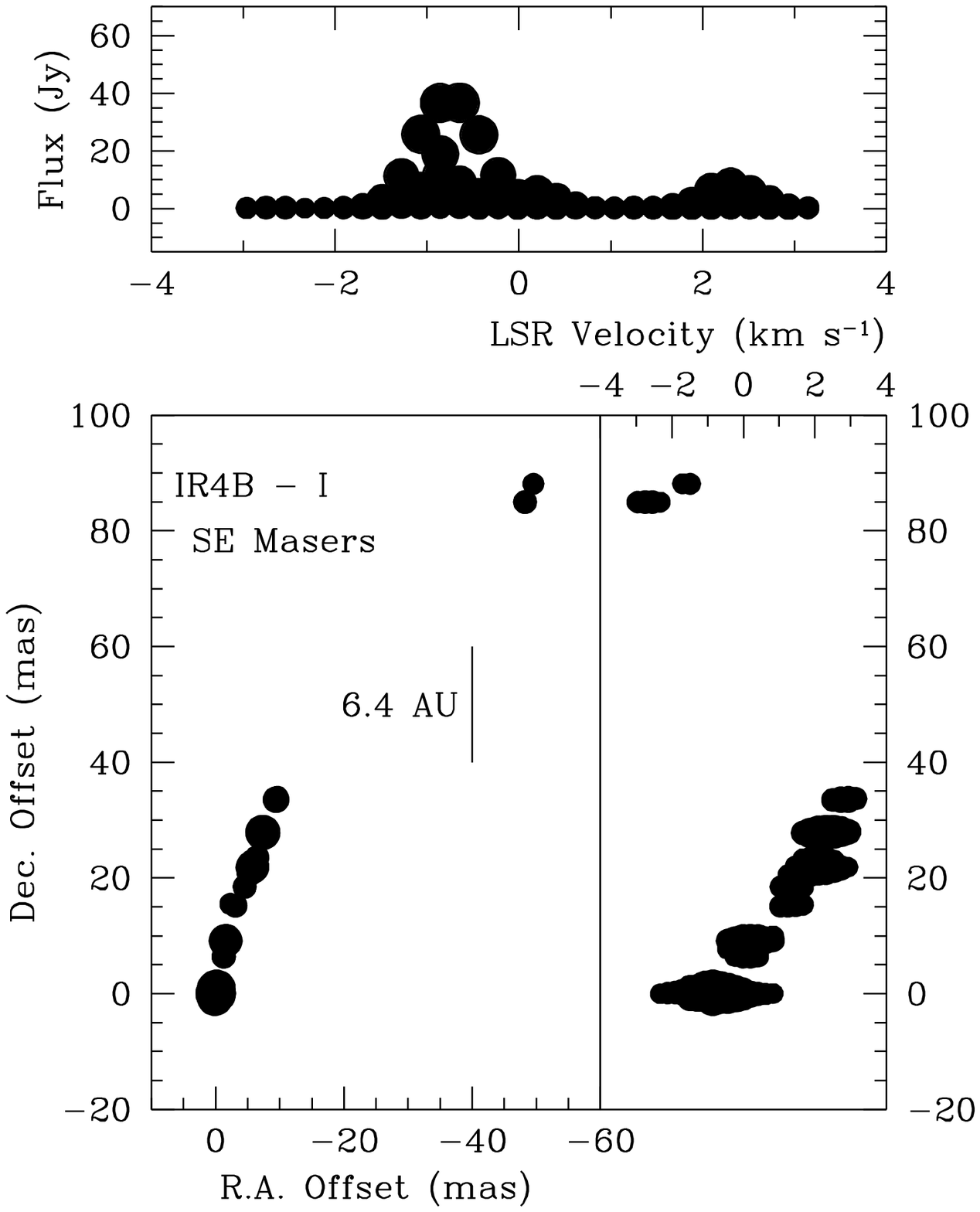}
\includegraphics[scale=0.43]{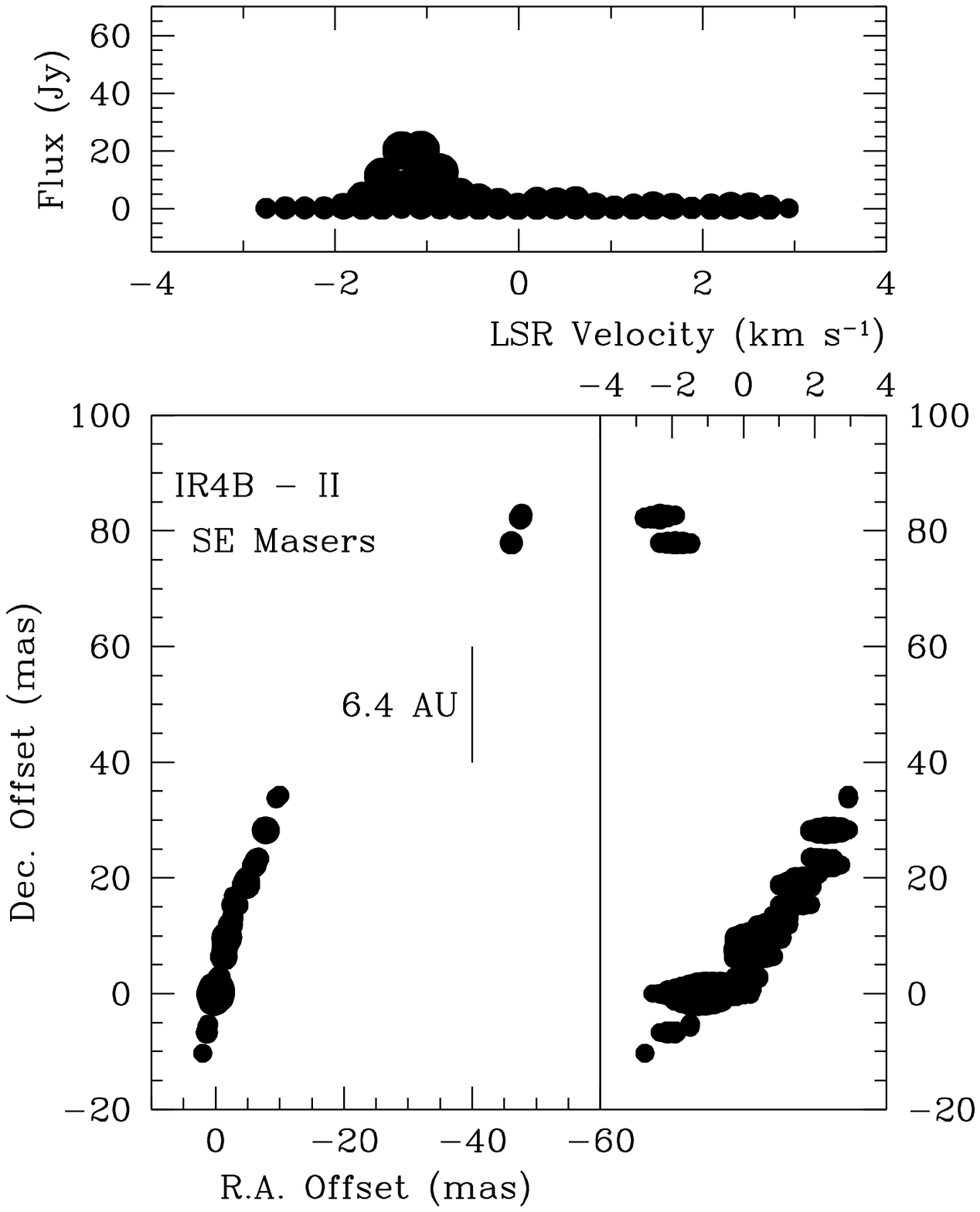}
\includegraphics[scale=0.43]{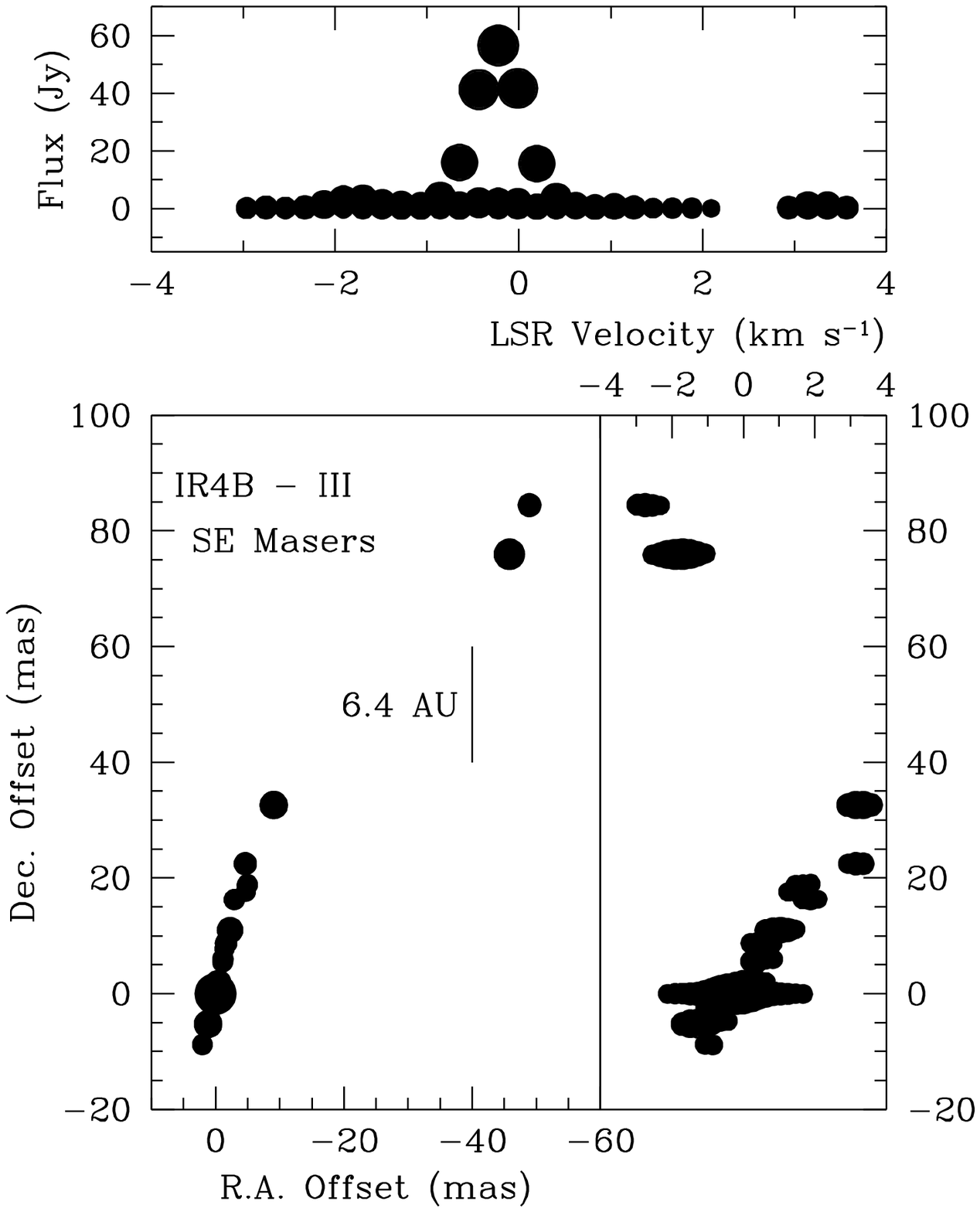}
\includegraphics[scale=0.43]{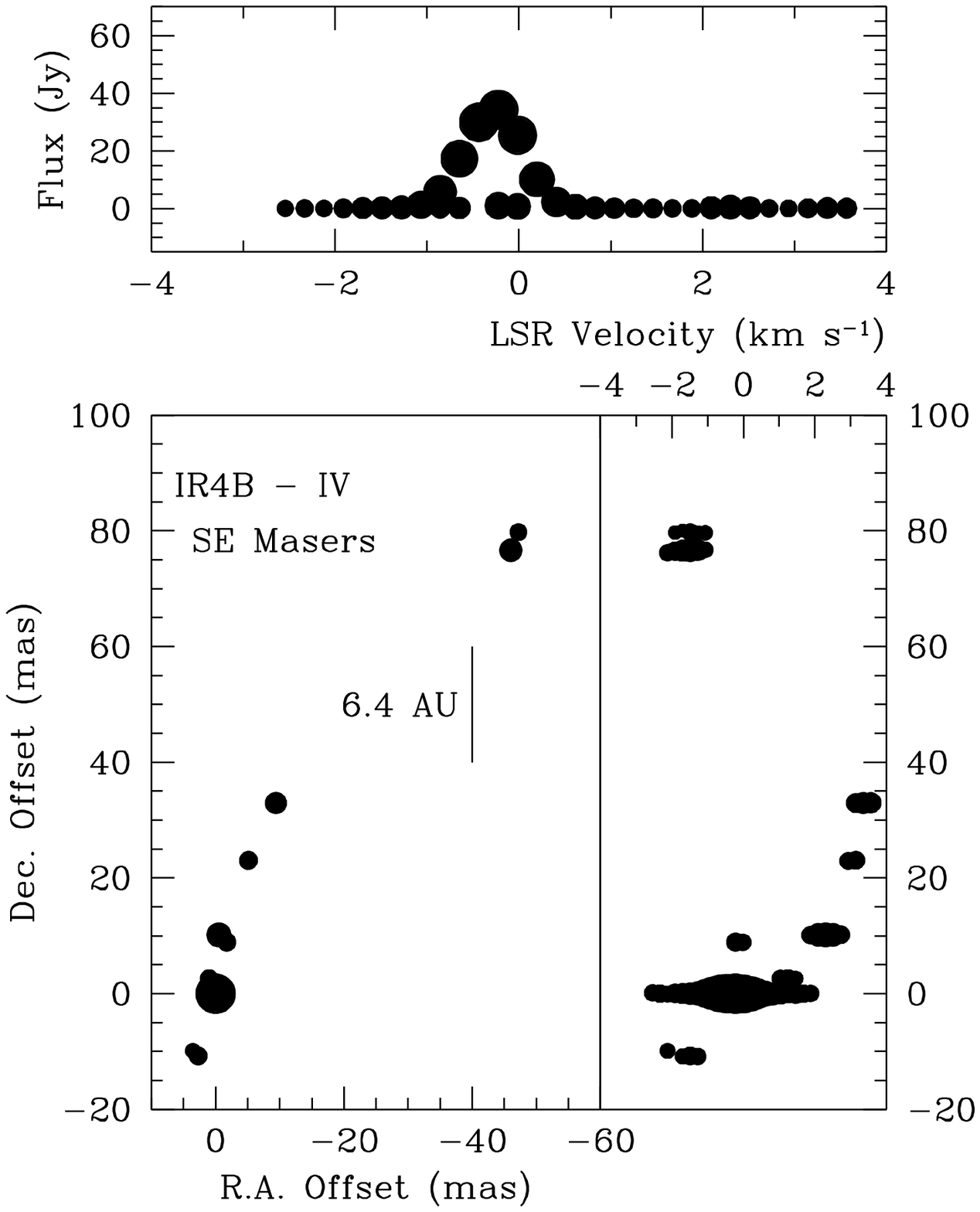}
\caption{
Spatial and kinematic distribution of the blueshifted masers southeast of IRAS~4BW for all four epochs.
In each panel, the top frame shows the velocity (color-coded) and the flux (symbol size) of the detected 
features, and the lower frame shows the spatial distribution and the velocity distribution as a function
of declination.  The uncertainty in the fitted position of each feature circle cannot be displayed at this 
scale, but in the worst case is of order 20 $\mu$arcseconds.
}
\notetoeditor{This four-panel color figure should appear on a single page and
each of the four panels should be legible.}
\end{figure}
\clearpage

\begin{figure}
\includegraphics[scale=0.43]{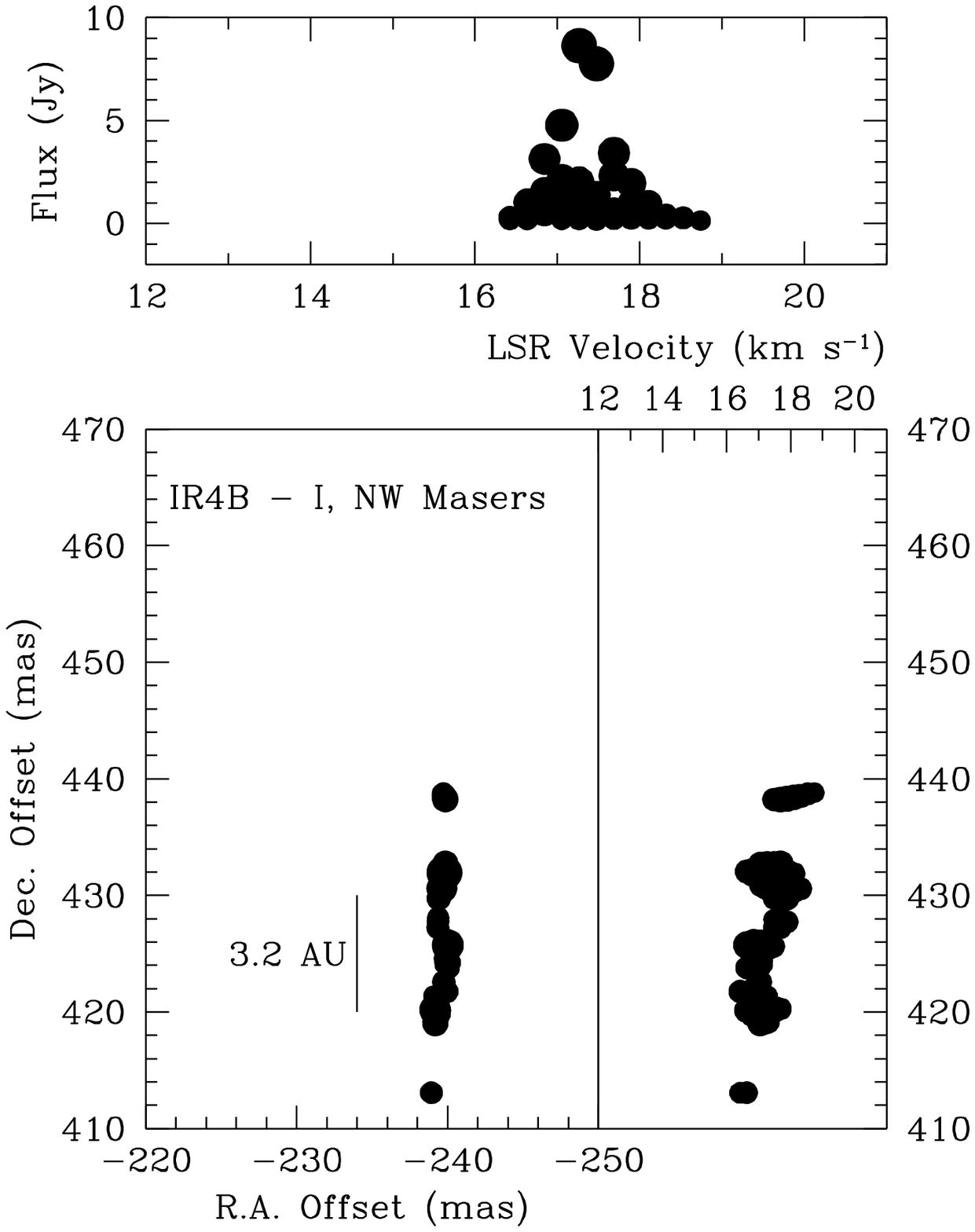}
\includegraphics[scale=0.43]{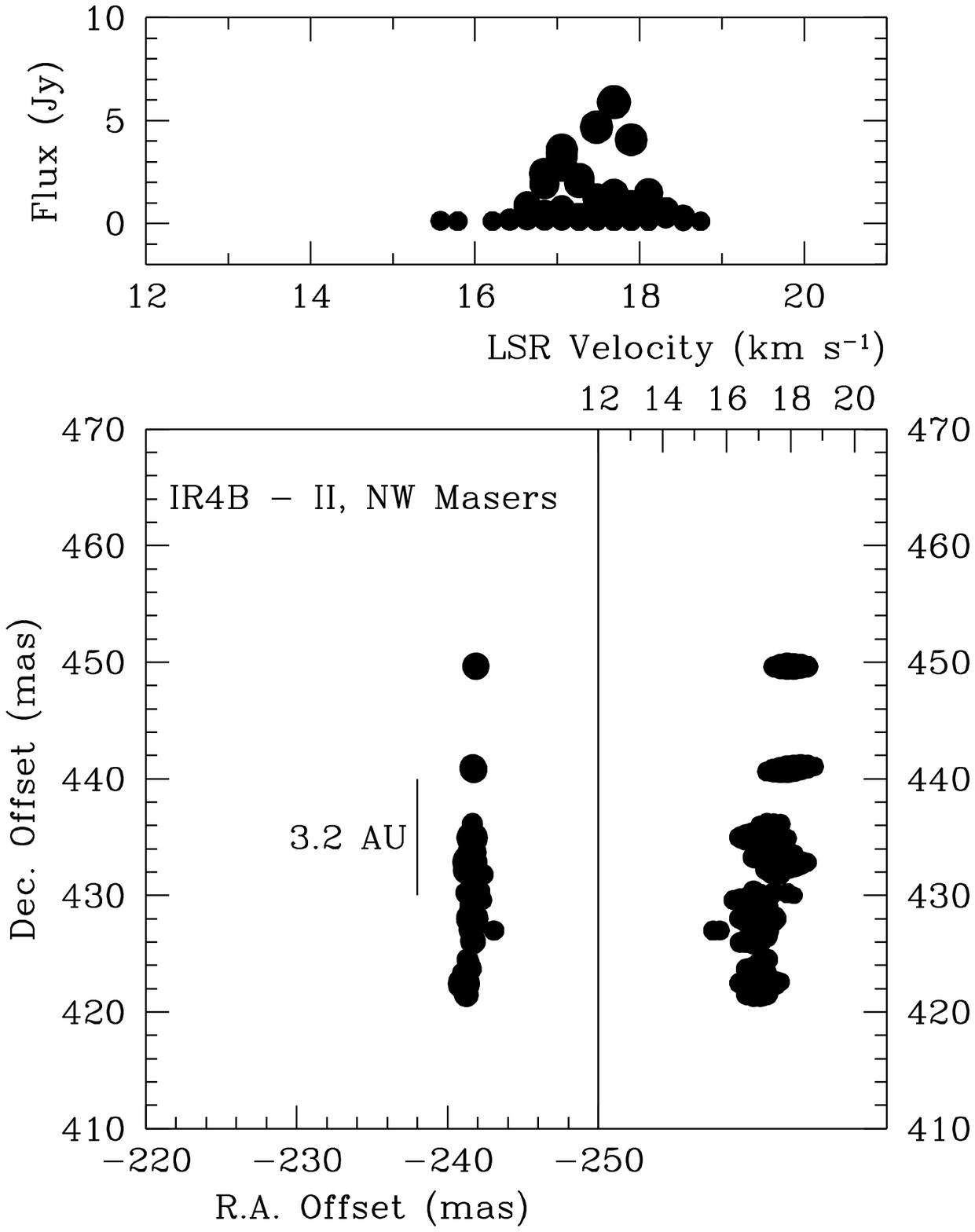}
\includegraphics[scale=0.43]{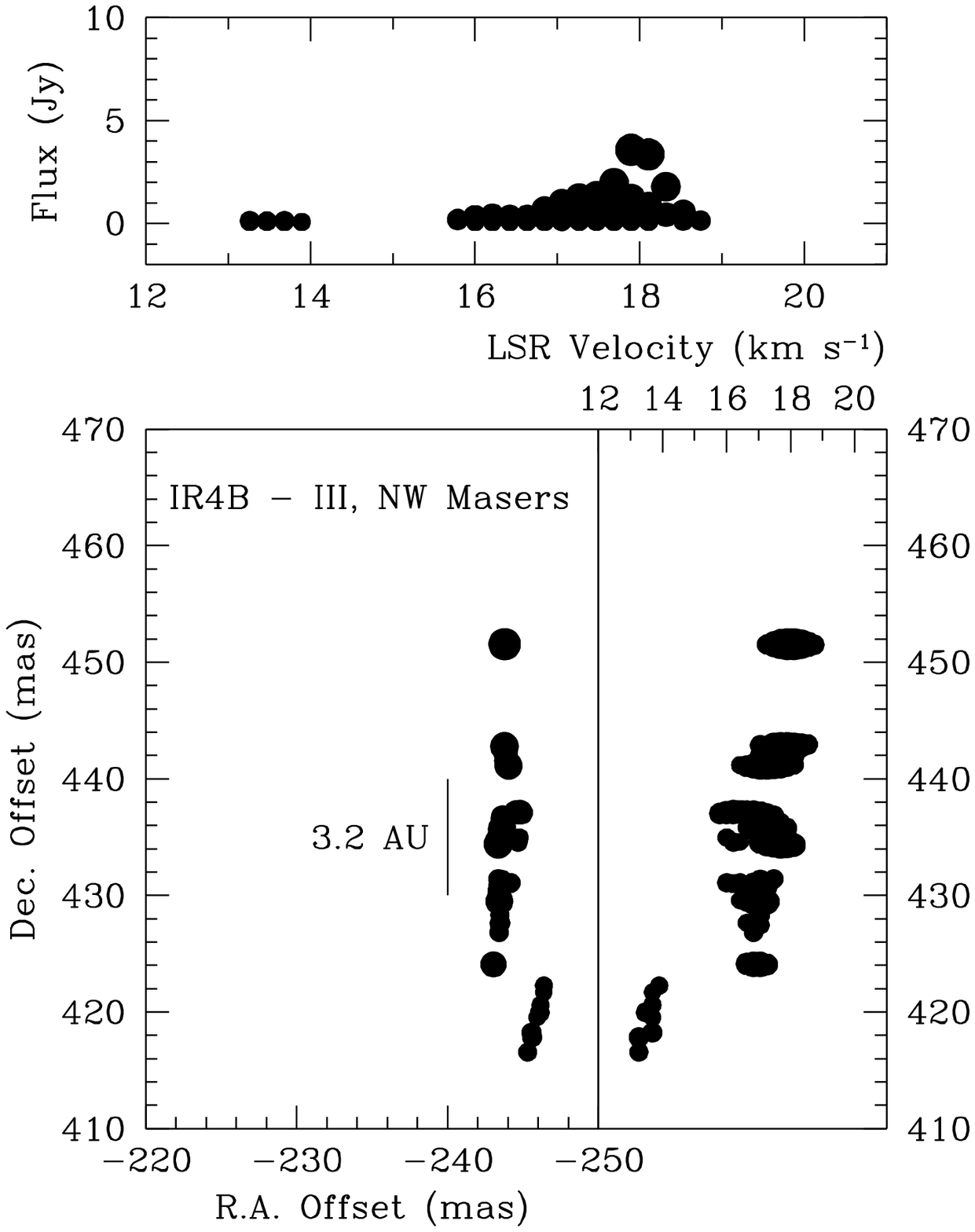}
\includegraphics[scale=0.43]{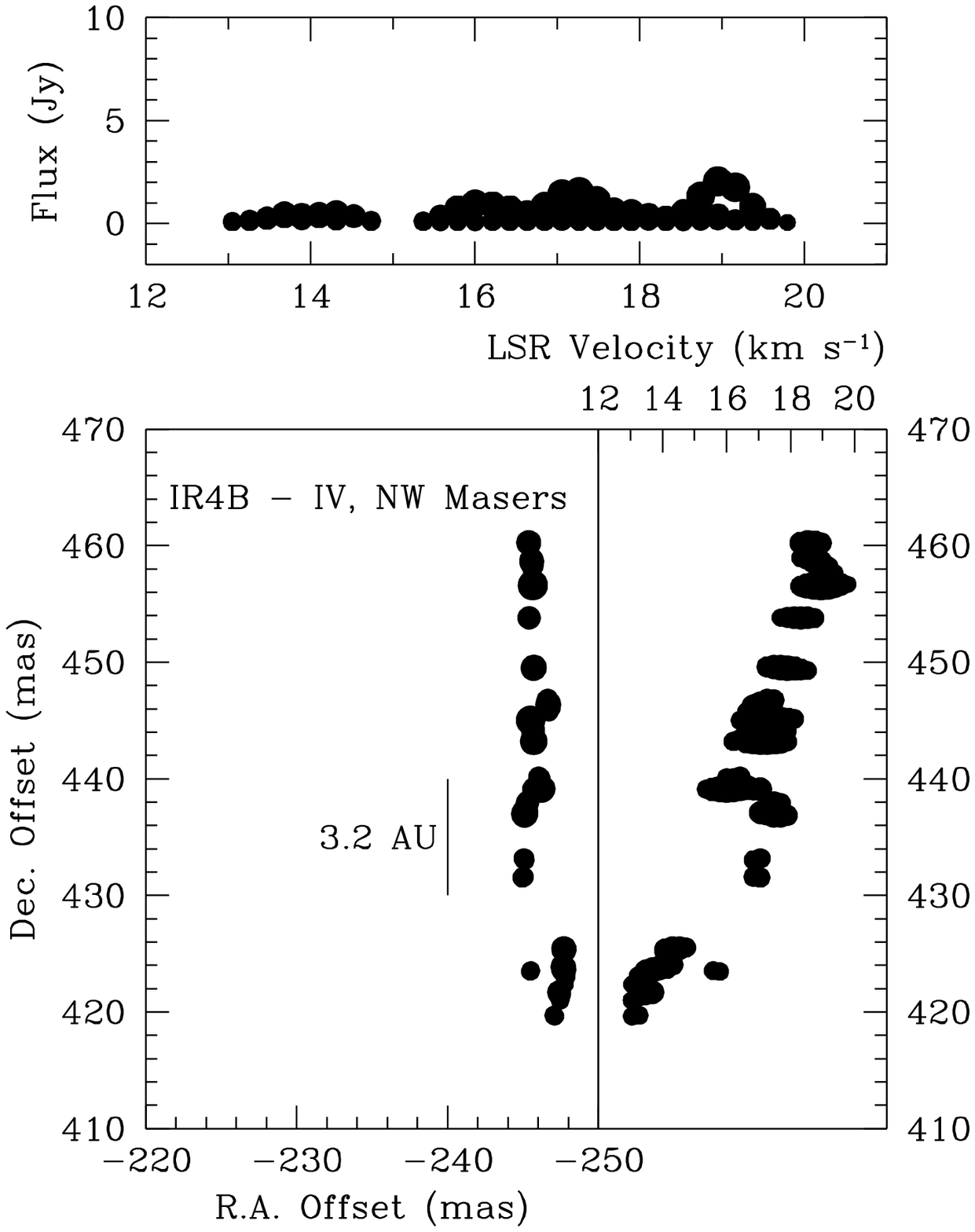}
\caption{
Spatial and kinematic distribution of the redshifted masers northwest of IRAS~4BW for all four epochs.
In each panel, the top frame shows the velocity (color-coded) and the flux (symbol size) of the detected 
features, and the lower frame shows the spatial distribution and the velocity distribution as a function
of declination.  The uncertainty in the fitted position of each feature circle cannot be displayed at this 
scale, but in the worst case is of order 20 $\mu$arcseconds.
}
\notetoeditor{This four-panel color figure should appear on a single page and
each of the four panels should be legible.}
\end{figure}
\clearpage
\begin{figure}
\includegraphics[scale=0.90]{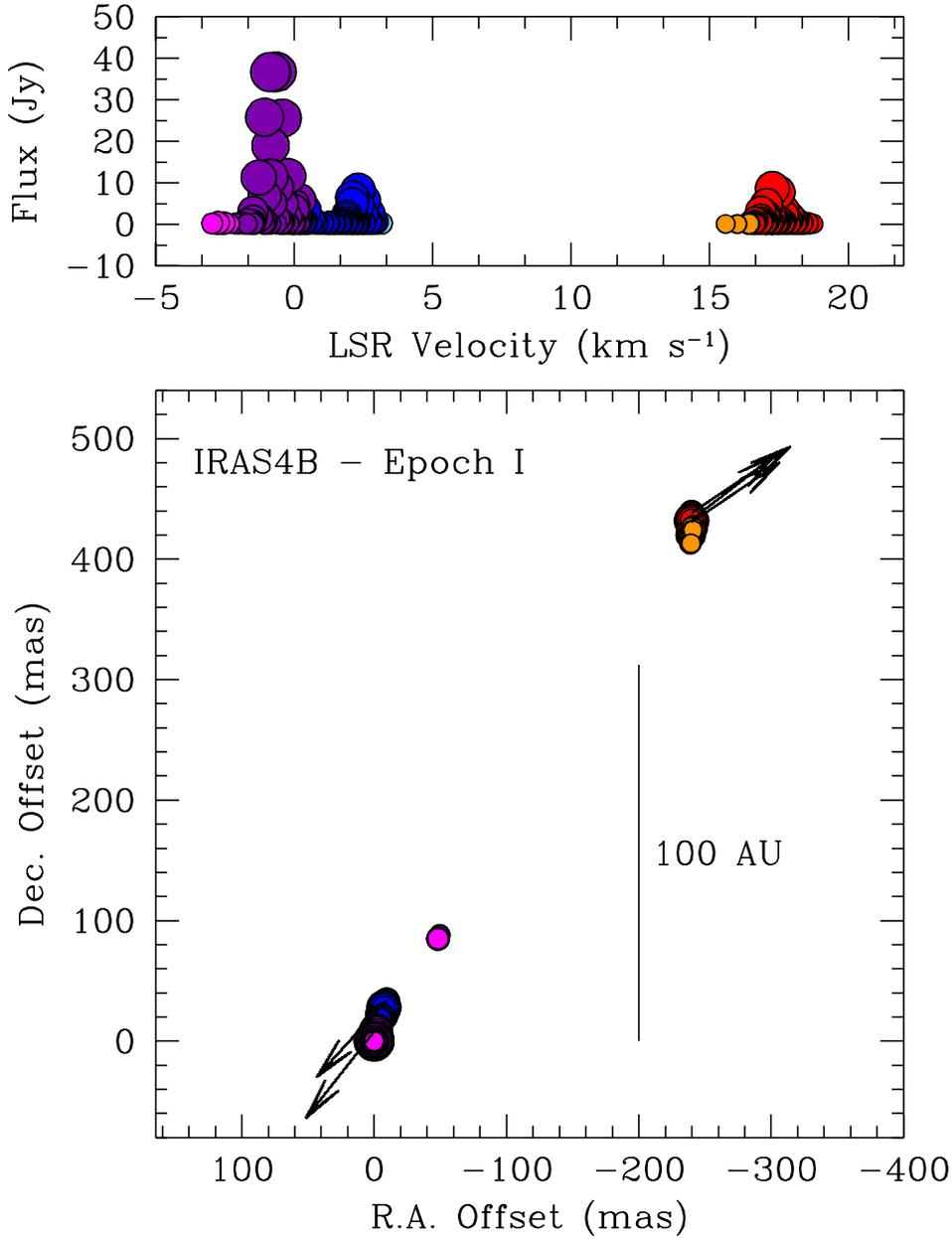}
\caption{
A plot of the proper motions for the five maser components present in all four epochs for IRAS~4BW relative
to their positions in Epoch 1 (see Table 4).  The length of the proper motion vectors are 2 \kms per plot unit.  
Note that the proper motions are not at the same position angle as a line connecting the two maser groups
(see text).
}
\end{figure}

\end{document}